\documentclass[12pt]{article}
\usepackage[margin=0.5in]{geometry}
\usepackage{amsmath,amssymb}
\usepackage{hyperref}
\hypersetup{
  pdftitle={Resource depletion accelerates rate learning but not composition learning in patch foraging},
  colorlinks=true,
  linkcolor=black,
  citecolor=black,
  urlcolor=blue
}
\usepackage{cite}
\usepackage{graphicx}
\usepackage{microtype}
\usepackage{lmodern}
\usepackage[T1]{fontenc}
\usepackage{titlesec}
\usepackage{authblk}
\usepackage{abstract}
\usepackage{booktabs}
\usepackage{array}
\usepackage{tabularx}
\newcolumntype{L}{>{\raggedright\arraybackslash}X}

\usepackage{pifont}
\newcommand{\mc}{\mathcal}
\newcommand{\Pp}{{\mc P}}
\usepackage{xcolor}

\titleformat{\section}{\large\bfseries\sffamily}{\thesection}{1em}{}
\titleformat{\subsection}{\normalsize\bfseries\sffamily}{\thesubsection}{1em}{}
\titleformat{\subsubsection}{\normalsize\itshape\sffamily}{\thesubsubsection}{1em}{}
\title{\sffamily\LARGE\bfseries Resource depletion accelerates rate learning but not composition learning in patch foraging}
\author[1,2]{Zachary P. Kilpatrick}
\author[3,4,5]{Ahmed El Hady}
\affil[1]{Department of Applied Mathematics, University of Colorado Boulder, Boulder, CO}
\affil[2]{Institute of Cognitive Science, University of Colorado Boulder, Boulder, CO}
\affil[3]{Department of Collective Behavior, Max Planck Institute of Animal Behavior, Konstanz, Germany}
\affil[4]{Department of Biology, University of Konstanz, Konstanz, Germany}
\affil[5]{Centre for Advanced Study of Collective Behavior, University of Konstanz, Konstanz, Germany}
\date{}

\begin{document}
\maketitle

\begin{abstract}
Foraging is a universal animal behavior that has increasingly attracted the interest of both experimentalists and theorists. Most prior models assume an animal knows the distribution of resources in its environment, but this structure must be learned as the animal explores its environment. Foraging can thus be regarded as a hierarchical inference problem. We develop a normative Bayesian account of an agent learning a patchy environment while exploiting it, and show that resource depletion shapes the levels of that hierarchy differently. Within a patch, depletion accelerates rate learning, since successive encounters occur at falling rates whose spacing pins down the initial rate. Across patches, composition learning, inferring the fraction of patches that are high yield, is slow, set by the number of patches sampled rather than the time spent in each, and unaffected by depletion once the rates are known. Reward-maximizing and information-seeking strategies therefore diverge, a forager resolving the composition underharvesting rich patches because learning it requires departures, most sharply early in exposure. When a fixed set of patches replenishes between visits, the reward-maximizing policy collapses onto a stable orbit over the high-yield patches, and the replenishment rate sets whether the forager maps the whole environment or locks onto a rich subset. Which departure rule maximizes intake is itself set by how variable the patches are, switching from counting prey to timing the gaps between them once richness varies by more than about a quarter. Learning the environment thus buys significant intake over learning a rule from reward alone, as long as its assumptions about depletion leave room for the truth.
\end{abstract}
\renewcommand{\abstractname}{Author summary}
\begin{abstract}
\noindent
{\footnotesize
Animals entering an unfamiliar landscape must both gather food and learn what kind of environment they face. We study this using a mathematical model of an animal moving among food patches and deciding when to keep feeding or leave.
Feeding depletes a patch, usually viewed only as a cost. We show it is also informative. As food is removed items arrive more slowly, and that pattern reveals how rich the patch was to begin with, so an animal learns patch quality faster where supply runs down than where it does not.
The benefit does not extend to every feature. Learning what fraction of patches are rich depends on how many patches are visited rather than how long the animal stays in each, so depletion neither helps nor harms it, and an animal seeking that information should leave even rich patches early to sample more locations.
Finally, the best rule for when to leave depends on how much patches differ. Where patches are alike an animal should count what it has eaten, and where they differ it should time the gap since its last find, because a long gap then signals a poor patch rather than an emptied one.}
\end{abstract}

\section{Introduction}

Throughout its life an animal must learn how to forage, adjusting its strategies as conditions change. A large body of work asks which decision strategies are optimal under which environmental conditions, often assuming the animal has full or partial knowledge of the underlying environmental structure~\cite{charnov1976optimal,stephens1986foraging,stephens1989variance,hughes1992optimizing}. Yet an animal new to an area does not know how resources are distributed and must infer that structure by exploring, thus improving later decisions~\cite{krebs1978test,kamil1982learning,regelmann1986learning,Eliassen_Jorgensen_Mangel_Giske_2009,constantino2015learning}. The marginal value theorem and its stochastic generalizations typically assume a forager knows its environment upon arrival~\cite{charnov1976optimal,oaten1977optimal}, an unlikely scenario. Rather, foraging animals appear to dynamically update their beliefs from resource encounters~\cite{webb2025foraging}. How animals learn the features of a foraging environment, and how that learning depends on the environment's structure, remains an open question. Such a model is central to survival, since it lets an animal generalize in ways a fixed policy cannot.

We treat learning as a second level of inference in patch foraging~\cite{yu2005uncertainty}. Classical patch-leaving theory assumes the relevant environmental statistics are known, and Bayesian extensions let a forager infer the type of the patch it currently occupies from within-patch encounters~\cite{charnov1976optimal,green1980bayesian,McNamara_1982,oaten1977optimal,olsson2006bayesian}. Here the forager must also learn the environment that generates those patches, the encounter rate of each patch type and the prevalence of each, which matters because patch-leaving decisions depend on expectations about the alternatives elsewhere as much as on the state of the patch at hand. Recent laboratory work shows that foragers learn the global opportunity cost from experience~\cite{constantino2015learning}, combine a prior over within-patch rate parameters with the reward times of the current patch~\cite{webb2025foraging}, and can overharvest through rational inference about latent patch structure rather than through a failure of optimization~\cite{harhen2023overharvesting}. These tasks present one patch type per session, whose rate decays with elapsed time, so there is no composition to infer and the timing of encounters adds nothing to the clock. We instead take a small, fixed number of patch types whose rates and prevalence are estimated jointly, with depletion driven by consumption rather than by time. Learning is then parametric inference over a few rates and frequencies rather than inference of an unbounded latent structure, consistent with evidence that animals combine prior experience with new observations to estimate environmental parameters~\cite{piet2018rats,tavoni2022human,valone2024probabilistic}.

Two broad framings are commonly used to study such learning, statistical inference using Bayesian updating and reinforcement learning that adjusts policies from experienced reward~\cite{Kolling_Akam_2017,sutton2018reinforcement,mobbs2018foraging}. Our Bayesian approach keeps an explicit posterior over the environmental parameters, while still providing analytically tractable learning dynamics, and we contrast with policy learning at the end. Within this framing an agent may seek to reduce its uncertainty about the environment~\cite{gottlieb2013information,kidd2015psychology,schwartenbeck2019computational}, rather than purely seeking reward. Reward-seeking and information-seeking objectives can prescribe different patterns of patch visitation, reflecting the older idea that information gathering is itself a primary drive rather than only a means to reward~\cite{inglis2001information}. Notably, information-maximizing policies trade mean reward for reliability in a way reward-maximizing ones do not~\cite{stephens1989variance,olsson2006foraging,barendregt2026information}.

Our main result is that depletion is a source of information, not only a cost. Each encounter lowers the rate at which the next arrives, so the spacing within a visit identifies the initial rate, and an agent in a depleting patch learns it faster than in a non-depleting one. The mechanism does not carry one level up, since composition, the fraction of each patch type, is set by the number of patches sampled rather than the encounters within them. Coupled to departure decisions, reward- and information-seeking strategies diverge most sharply early in experience, and even among information-seeking strategies natural measures of learning can favor movement in opposite directions. When patches replenish the same split reappears under a single parameter, since fast recovery leaves a returning forager well informed about a few patches and ignorant of the rest. Which departure rule is best is set by the variability of the patches rather than their richness, switching from a count to a timer as richness spreads, and an agent representing the environment outperforms one learning a rule from reward, though only while its model of depletion is roughly right. Depletion thus reveals the hidden structure of the environment, movement determines which parts of it can be learned, and modeling it outperforms learning a rule from reward alone as long as the model is not too badly misspecified.

\section{Model}

We model an animal (agent) searching a large arena (environment) containing patches of food whose spatial scale is smaller than the distance between them~\cite{pyke2015understanding}. Such geometry favors a strategy in which the agent enters a patch, consumes food within it, and departs in search of another~\cite{charnov1976optimal,searle2005should}. Our prior work assumed the distribution of patch types $p(\lambda)$ was known, so the agent needed only to infer the type of the patch it occupied~\cite{kilpatrick2021uncertainty}. Animals entering a new environment do not know $p(\lambda)$~\cite{kamil1982learning,regelmann1986learning,Eliassen_Jorgensen_Mangel_Giske_2009,constantino2015learning}, and must estimate it from their encounters. This is a second, higher level of inference, in which the agent updates a posterior over the initial rate, $p(\lambda | x_{1:n})$, across the past $n$ visits, taking each newly entered patch as undepleted.

\begin{figure}[t!]
    \centering
    \includegraphics[width=17cm]{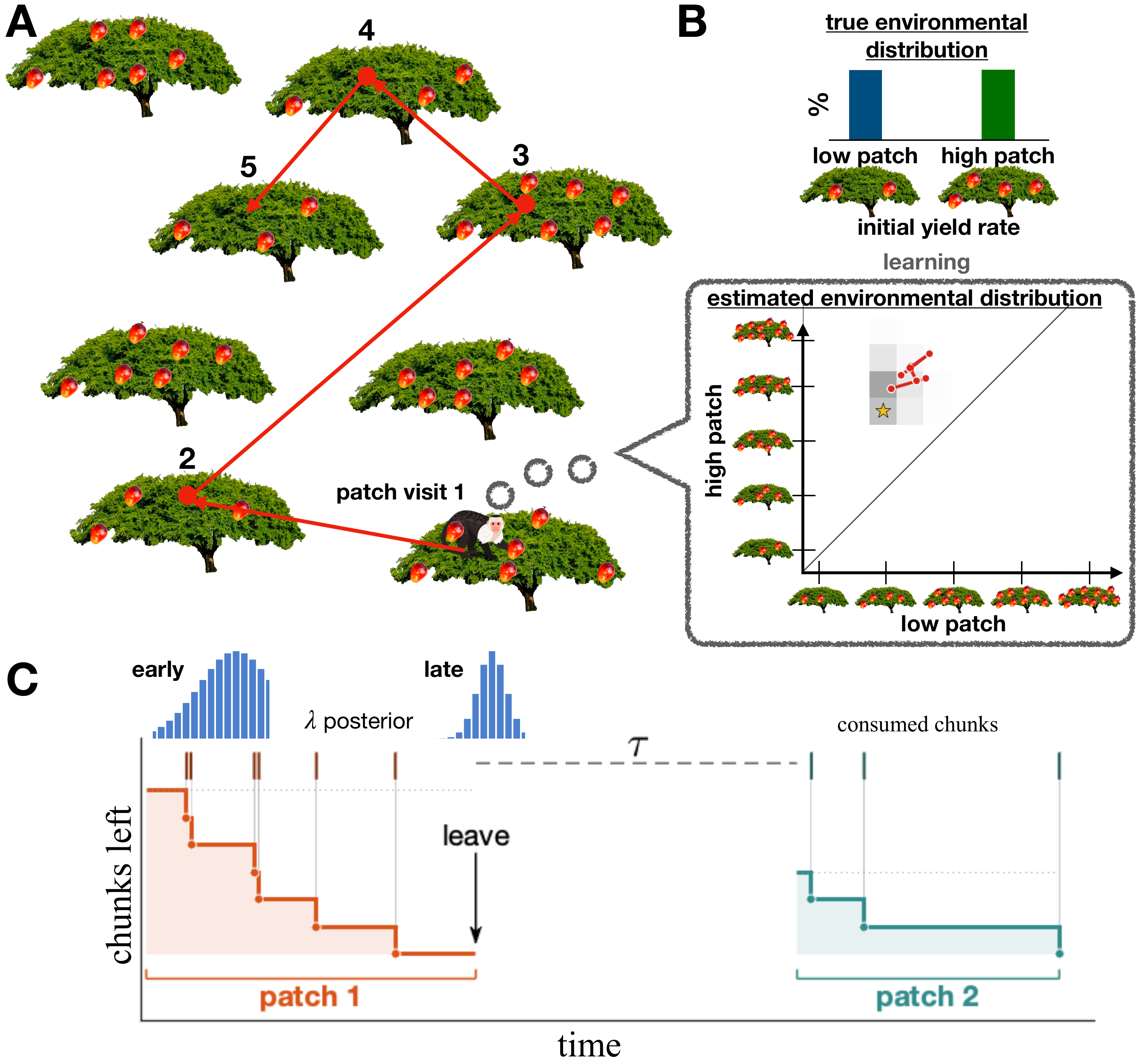}
    \caption{{\bf The two levels of inference in patch foraging.} {\bf A.}~An animal moves among trees (patches), each with its own initial yield rate setting the rate at which it encounters fruit (chunks) on entry, paying a travel delay $\tau$ between them. Numbers give the visit order of one realization, traced in red. {\bf B.}~Top, the true environment, a fraction of low- and high-yield trees each with its own rate. Bottom, the agent's estimate of the two rates, refining across visits (path) against the true pair (star). Because each chunk lowers the rate by $\rho$, the rates lie on the lattice $m \rho$. {\bf C.}~Within a single visit. Ticks mark chunk encounters and the staircase the chunks left, and the posterior over the patch's initial count sharpens from early to late. The animal then departs and pays $\tau$ to reach a fresh patch.}
    \label{fig1:scheme}
\end{figure}

{\em Encounters and depletion.} Patch type $k$ carries an initial encounter rate $\lambda_k$, so the time to the first {\em food chunk} after entry is exponential, $p(t_1 | \lambda_k) = \lambda_k e^{- \lambda_k t_1}$. Each chunk found lowers the rate by $\rho$, giving $\lambda(t) = \lambda_k - \rho K(t)$ for $K(t)$ chunks found so far, so the interval between the $K$th and $(K\!+\!1)$th encounter is drawn from
\begin{align}
    p(t_{K+1} | \lambda_k) = (\lambda_k - \rho K) e^{- (\lambda_k - \rho K)t_{K+1}}. \label{interval}
\end{align}
The decrement $\rho$ is small when a patch holds many chunks and large when it holds few. A patch of type $k$ holds $m_0^k = \lambda_k / \rho$ chunks in total, and rates lie on the lattice $m \rho$, so we work in terms of the patch richness $m_0$ rather than $\lambda_k$ throughout. In the limit $\rho \to 0$ the rate is fixed within a visit, the {\em non-depleting} case, a good approximation when a patch holds many chunks and the agent departs before depleting it appreciably. On departing, the agent pays a travel delay $\tau$, which may be drawn from a distribution, to reach another patch whose initial rate is drawn from $p(\lambda)$ (Fig.~\ref{fig1:scheme}).

{\em Reward rate.} The efficiency of a foraging policy $\Pp$, a rule mapping the encounter history within a patch to a departure decision, is measured by the reward rate. Averaging the resources gained and the times spent over many visits (Methods) gives
\begin{align}
    R(\Pp) = \frac{\int_0^{\infty} \langle r | \lambda \rangle_{\Pp} \cdot p(\lambda) d \lambda}{\int_0^{\infty} \langle T | \lambda \rangle_{\Pp} \cdot p(\lambda) d \lambda + \langle \tau \rangle_{\Pp}}. \label{energygain}
\end{align}
Departure strategies are then found by relating reward intake and patch visit statistics to departure rules and optimizing. Because the rate falls only at chunk encounters, the reward the agent stands to gain changes only at encounters, but its belief moves continuously, since the intervals without food are themselves evidence and lower the posterior over the patch's richness faster the richer the patch is. We restrict throughout to threshold policies, which depart once the belief about the occupied patch crosses a fixed level, whether an encounter or a lengthening dead time (with no food) carries it there.

Requiring the agent to infer $p(\lambda)$ over time complicates how a policy should be chosen. Since inferring full densities is costly, we consider a small finite number of patch types $N_T$, so that $p(\lambda) = \sum_{k=1}^{N_T} p_k \delta (\lambda - \lambda_k)$ with $\sum_{k=1}^{N_T} p_k = 1$. This frames learning as a {\em parametric inference} problem, a skill identified in humans and other animals~\cite{tavoni2022human,piet2018rats}, and it suggests an agent may care not only about reward rate but also about reducing its uncertainty over $p(\lambda)$.

{\em Information rate.} An agent may thus be primarily information seeking rather than reward seeking, especially in unfamiliar environments~\cite{stephens1989variance,olsson2006foraging}. An alternative measure of a policy $\Pp$ is its information gain rate, the reduction in entropy divided by time~\cite{lindley1956measure,mackay1992information},
\begin{align}
    IGR_n = \frac{H_0 - H_n}{\sum_{j=1}^n [T_j + \tau_j]}, \label{infrate}
\end{align}
where $H_n = - \int_{\Theta} p(\theta| x_n) \log p(\theta | x_n) d \theta$ is the entropy of the distribution over the environment's parameters $\theta$ given observations $x_n$ from $n$ visits. For $N_T = 2$, $\theta = \{ \lambda_H, \lambda_L, p_H \}$ collects the two arrival rates and the fraction $p_H$ of high patches.

A key question is how reward- and information-seeking strategies differ in patch visitation. Most of this work concerns agents initially uncertain about the arrival rate $\lambda_k$ in patches of type $k$, which they can resolve over a few long visits, though we also treat inference of the fractions $p_k$, for which uncertainty reduction can require many. We measure learning by the mean squared error (MSE) of the rate estimate, and where a departure rule needs a stopping criterion we threshold the posterior standard deviation instead, which tracks the same quantity but is monotone in accumulated evidence. We build from homogeneous to binary environments, where learning requires visiting more than one patch type.

\section{Results}

\subsection{Learning patch yield rates}
\label{sec:rates}

We begin with the rates themselves, first in a {\em homogeneous} environment where every patch shares a single rate $\lambda_1$, so $p(\lambda) = \delta(\lambda - \lambda_1)$ and the agent updates one univariate posterior carried from patch to patch, then in a {\em binary} environment with two rates $\lambda_H > \lambda_L$, where learning requires sampling both types. In each case we contrast non-depleting patches, $\rho \to 0$, with depleting ones.

\begin{figure}[t!]
    \centering
    \includegraphics[width=16cm]{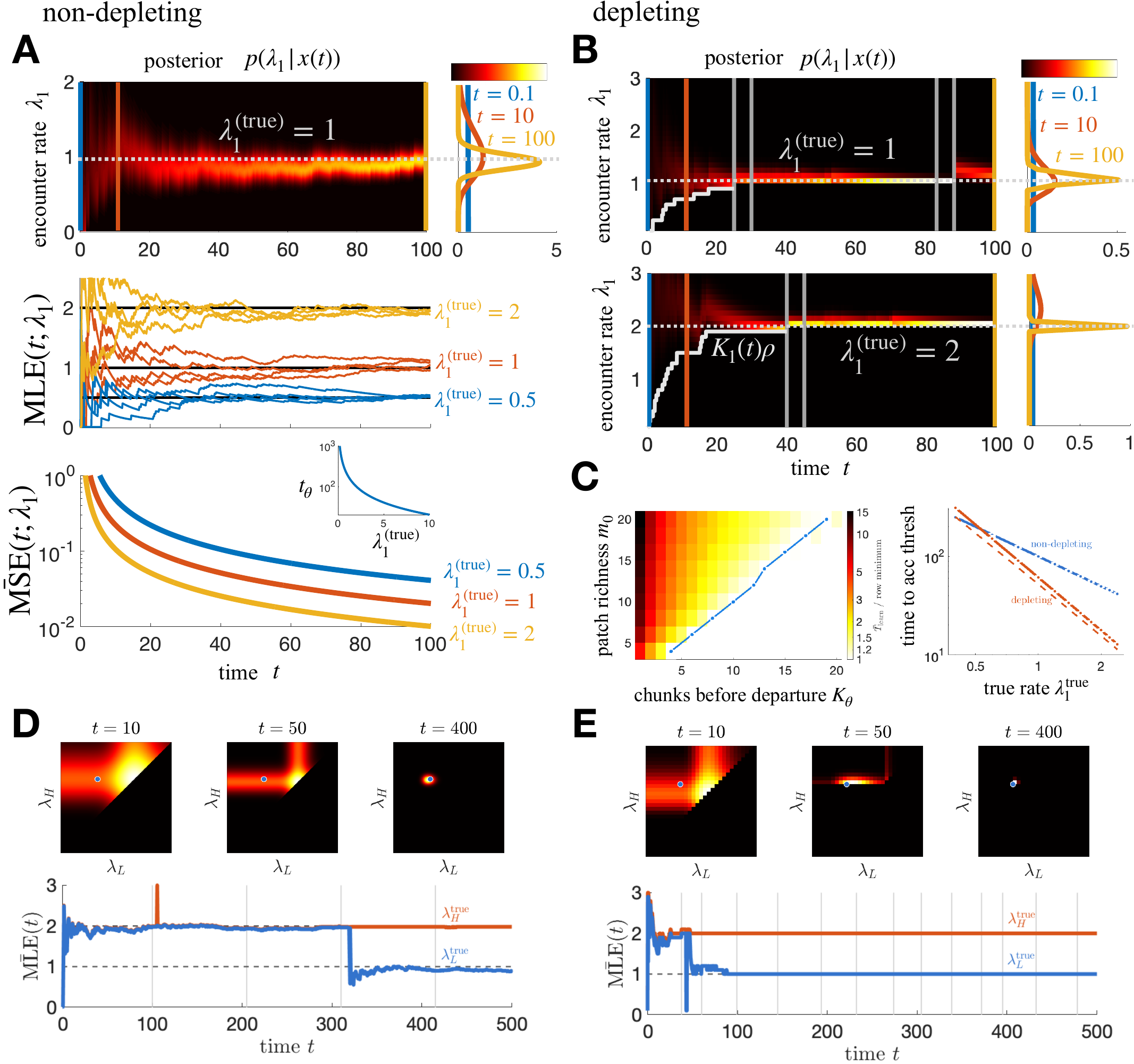}
 \caption{\textbf{Patch yield rates are learned faster in depleting than non-depleting environments, in homogeneous and binary environments alike.} \textbf{A.}~Homogeneous, non-depleting, $\lambda_1^{\rm true} = 1$ unless indicated. Top, the posterior $p(\lambda_1|x(t))$ sharpens over time, with line cuts at right. Middle, MLE traces at three true rates. Bottom, the averaged MSE of Eq.~(\ref{msevthomno}) decays faster for higher $\lambda_1^{\rm true}$, and inset, the time $t_\theta$ to fixed accuracy falls with it. \textbf{B.}~Homogeneous, depleting, at $\lambda_1^{\rm true} = 1$ (top) and $2$ (bottom). The posterior peaks faster than in A because successive intervals are drawn at falling rates. White staircase marks $K_1(t)\rho$. \textbf{C.}~Left, mean time to learn to fixed posterior accuracy over the consumed-count threshold $K_\theta$ and richness plane, each row normalized by its own minimum, so the color is the penalty for departing early, with the per-richness optimum traced. Leaving early costs up to fifteenfold, and the optimum sits at $K_\theta = m_0$ or one chunk short of it. Right, time to accuracy in the two environments from the Cram\'er--Rao bound on Eqs.~(\ref{depoisfi}) and~(\ref{poissfi}). The depleting advantage grows with rate, from parity near $\lambda_1^{\rm true} = 0.5$ to close to threefold at $2$; dashed, the same agent at $\tau = 0$. \textbf{D.}~Binary, non-depleting, $\lambda_H^{\rm true} = 2$ and $\lambda_L^{\rm true} = 1$. Top, the joint posterior at $t=10,50,400$ concentrates on the true pair (dot). Bottom, MLE traces with departures in grey. The occupied type moves while the under-sampled type holds fixed, and $\lambda_L$ resolves only once a low patch is entered. \textbf{E.}~Binary, depleting, same rates and times as D. Both rates lock about three times sooner. The staircase edge at $t=10$ and $50$ is the bound $\lambda \geq K\rho$, a consequence of the falling rate rather than a separate source of information. Low rates are never fully excluded, since the agent may always be in a high patch. Here $\rho = 0.1$ in the depleting panels.}
    \label{fig2_rates}
\end{figure}

\subsubsection{Homogeneous environments}

{\em Non-depleting.} In a non-depleting environment $\lambda(t) \equiv \lambda_1$, an agent seeking only to learn $\lambda_1$ as fast as possible should remain in a single patch indefinitely. All patches are the same, so moving gains no information and transit only wastes time, and the agent carries its posterior forward but adds nothing to it. Given encounters $x(t) = \sum_{j=1}^{K(t)} \delta (t - t_j)$, the observer combines the elapsed time $t$, the count $K(t) = \int_0^t x(s) ds$ and a prior $p_0(\lambda_1)$ through Bayes rule,
\begin{align}
    p(\lambda_1|x(t)) &= \frac{1}{p(K(t))} p(K(t)|\lambda_1) p_0(\lambda_1) \propto (\lambda_1 t)^{K(t)} e^{- \lambda_1 t} p_0(\lambda_1).  \label{nondeplearn}
\end{align}
The maximum likelihood estimate (MLE) is consistent, reducing to $\hat{\lambda}_1 = K(t)/t$ under a flat improper prior and to $K(t)/(t+t_0)$ under an exponential prior $p_0(\lambda_1) = t_0 e^{-t_0\lambda_1}$, the latter carrying a bias that decays algebraically (Methods). Both encounters and the intervals without food narrow the posterior, and the MLE moves toward the true rate (Fig.~\ref{fig2_rates}A).

Averaging the posterior over realizations of the encounter process gives a closed form for the mean normalized MSE,
\begin{align}
    \bar{\rm MSE}(t; \lambda_1^{\rm true}) = \int_0^{\infty} \frac{(\lambda_1 - \lambda_1^{\rm true})^2}{(\lambda_1^{\rm true})^2} \bar{p}(\lambda_1 | x(t)) d \lambda_1, \label{msevthomno}
\end{align}
whose kernel involves a modified Bessel function of the first kind and rescales under $\lambda_1^{\rm true} x = \lambda_1$, $s = \lambda_1^{\rm true} t$ (Methods). The decay rate of the normalized mean MSE therefore increases linearly in $\lambda_1^{\rm true}$, so the time $t_\theta$ to a threshold level of accuracy scales inversely with the true rate (Fig.~\ref{fig2_rates}A, inset). The observer learns the arrival rate more quickly in richer environments, since more frequent encounters resolve the rate sooner.

Biases in $p_0(\lambda_1)$ persist here, since they decay only algebraically, and most when the true rate is one the prior considers uncommon. Information gain grows only logarithmically in time at an asymptotic rate independent of the true rate, since higher rates deliver more encounters but are correspondingly harder to pin down (Methods). Staying is reward-optimal, $R_{\rm non\text{-}dep} = \lambda_1$. This is the baseline against which we compare depleting environments.

{\em Depleting.} In depleting environments the learning problem changes in two ways. First, the intervals within a visit are drawn successively at $\lambda_1, \lambda_1 - \rho, \lambda_1 - 2\rho$ and so on, and the spacing of this decelerating sequence identifies $\lambda_1$ far more sharply than a constant-rate stream of the same length~\cite{iwasa1981prey}. Each encounter also implies a lower bound on the initial rate, $\lambda_1 \geq K(t)\rho$, but that bound follows from the same falling rate rather than adding to it (Methods). Second, departing to a fresh patch resets the rate to $\lambda_1^{\rm true}$, so the encounter dynamics of a dwelt-in patch and a fresh one are genuinely different alternatives. The same within-visit information also overwrites the prior within a few encounters, so priors that differ by a factor of $21$ in rate-estimate error after a single patch differ by at most $4.4$ percent in long-run reward rate (S1~Text).

Bayes rule gives the posterior within a single patch,
\begin{align*}
    p(\lambda_1 | x(t)) & \propto \frac{m_0! e^{- \lambda_1 t} p_0(\lambda_1)}{(m_0 - K(t))!} ,
\end{align*}
for $\lambda_1 \geq K(t) \rho$ and zero otherwise, where $m_0 = \lambda_1 / \rho$ is the initial count associated with rate $\lambda_1$, so $\lambda_1$ is an integer multiple of $\rho$ and the prior and posterior are discrete. The distribution narrows as the patch empties (Fig.~\ref{fig2_rates}B). Across $n$ visits the forager accumulates evidence as a product,
\begin{align}
    p(\lambda_1 | x(t)) \propto &  \prod_{j=1}^{n} \frac{m_0! e^{- \lambda_1  t_j}}{(m_0 - K_j(t_j))!} \cdot p_0(\lambda_1), \label{homlearndepj}
\end{align}
with $t_j$ the time spent in the $j$th patch and $K_j(t_j)$ the chunks encountered there.

One may ask why, given the travel cost $\tau$, it is not simply optimal to remain in a single patch indefinitely. The answer depends on which measure of learning we adopt, and in the homogeneous case two natural measures disagree. Under an entropy-rate measure, no-exit is optimal. Every patch shares the same rate, so a fresh patch furnishes no new parameter, only further samples of $\lambda_1$, and the depleted patch keeps lowering posterior entropy through its slowing encounters and through the informative dead time between them. Departing forfeits this cheap evidence and pays $\tau$ to restart from $K=0$, so the entropy gain rate from staying exceeds that from going at every chunk count, even for small $\tau$.

The MSE-threshold measure suggests a different strategy, since depletion makes the slow tail of an exhausted patch a poor use of time relative to the faster early encounters of a fresh one. Under this measure the time to threshold is minimized when the chunks consumed reach the MLE of the initial count, $K_{\theta}^{\rm opt} = {\rm MLE}(m_0)$, writing $K_\theta$ for a threshold on the consumed count. The time to threshold falls steeply in $K_\theta$, by up to fifteenfold, and is minimized at $K_\theta = m_0$ up to $m_0 = 16$ and one chunk short of it beyond, so the best policy is to consume essentially the whole patch and then depart (Fig.~\ref{fig2_rates}C, left). The two measures disagree over the emptied patch's dead time, which a rate objective takes because it is free and a stopping-time objective refuses because it is slow. The advantage over a non-depleting environment follows from the Cram\'er--Rao bound on Eqs.~(\ref{depoisfi}) and~(\ref{poissfi}), with no simulation (Fig.~\ref{fig2_rates}C, right). It grows with rate, from parity near $\lambda_1^{\rm true} = 0.5$ to close to threefold at $2$, because the falling rates within a visit identify $\lambda_1$ more sharply, and below the crossover the cost of replacing an exhausted patch outweighs that gain.

\subsubsection{Binary environments}
\label{sec:binary}

A forager learning two rates $\lambda_H>\lambda_L$ must visit more than one patch, and each successive visit raises the chance that both types have been sampled.

{\em Non-depleting.} In a single patch $j$ with a flat prior, the encounters $x_j(t)$ yield $p(\lambda | x_j) \propto (\lambda t_j)^{K_j(t_j)} e^{- \lambda t_j} \equiv Q(\lambda; t_j, K_j)$ as in Eq.~(\ref{nondeplearn}). Given $n$ visits and a flat prior over the pair, the forager combines evidence by marginalizing over the unknown type of each patch,
\begin{align*}
    p(\lambda_H, \lambda_L |x(t)) &\propto \prod_{j=1}^n \left[ p_H Q(\lambda_H; t_j, K_j) + p_L Q(\lambda_L; t_j, K_j) \right], \ \ \ \ \lambda_H > \lambda_L,
\end{align*}
where $p_k$ is the (here known) fraction of type $k$. The posterior refines around the rate of the type currently occupied, and in the MLE time series the estimate for the type {\em not} occupied holds roughly fixed while that of the occupied type moves (Fig.~\ref{fig2_rates}D).

{\em Depleting.} The single-patch posterior follows Eq.~(\ref{homlearndepj}), giving $Q_{\rho}^j(\lambda) \propto [(\lambda/\rho)!/(\lambda/\rho-K_j)!]\,e^{-\lambda t_j}$ with a flat prior, and marginalizing over the visited type gives
\begin{align*}
    p(\lambda_H, \lambda_L | x(t)) &\propto \prod_{j=1}^n \left[ p_H Q_{\rho}^j(\lambda_H) + p_L Q_{\rho}^j(\lambda_L) \right], \ \ \ \ \lambda_H > \lambda_L.
\end{align*}
As in the homogeneous case the posterior first refines over the type visited first (Fig.~\ref{fig2_rates}E). Unlike the homogeneous case, low rates $\lambda_L < K\rho$ are never fully excluded, since the agent may always be in a high patch, so the high-rate MLE sharpens quickly while the low-rate MLE fluctuates until a low patch is entered (Fig.~\ref{fig2_rates}E).

This asymmetry is what makes departure informative in heterogeneous environments, in contrast to the homogeneous case where no-exit was entropy-rate optimal. Suppose the agent has visited only high patches and seeks the unobserved $\lambda_L$, with marginal posterior entropy $H_{\lambda_L}$. Writing the marginal information gain rates from staying over a further increment and from departing as
\begin{align}
    IGR_{\lambda_L}^{\rm stay} = \frac{H_{\lambda_L}^{\rm curr} - \langle H_{\lambda_L}^{\rm stay} \rangle}{\langle \Delta t \rangle}, \qquad
    IGR_{\lambda_L}^{\rm depart} = \frac{H_{\lambda_L}^{\rm curr} - \langle H_{\lambda_L}^{\rm next} \rangle}{\tau + \langle t_{\rm next} \rangle}, \label{igrstay}
\end{align}
with $\langle t_{\rm next} \rangle = p_H / \lambda_H + p_L / \lambda_L$ the mean time to the first chunk in a fresh patch under the type prior, the stay numerator is $O(p_L e^{-(\lambda_H-\lambda_L)\Delta t})$, exponentially small in the rate gap, while the departure numerator is $O(p_L)$, bounded away from zero whenever the fresh patch can be the under-sampled type (Methods). Once the current type is resolved the departing rate exceeds the staying rate and the agent should leave. Depletion does more than speed within-patch inference. By making the under-sampled rate's marginal uncertainty the binding constraint, it turns departure into the faster route to a complete description of the environment.

\subsection{Learning the environmental composition}
\label{sec:composition}

Rate inference is fast because every interval within a visit is drawn at a different rate, many bits per patch. Learning the {\em fraction} $p_H$ of high patches is slower and structurally different, since each patch contributes at most one categorical observation, its own type, however richly it is sampled within. We take the rates as known, resolved quickly as above, and ask how the agent infers $p_H$ from the sequence of types.

\begin{figure}[t!]
\begin{center} \includegraphics[width=12cm]{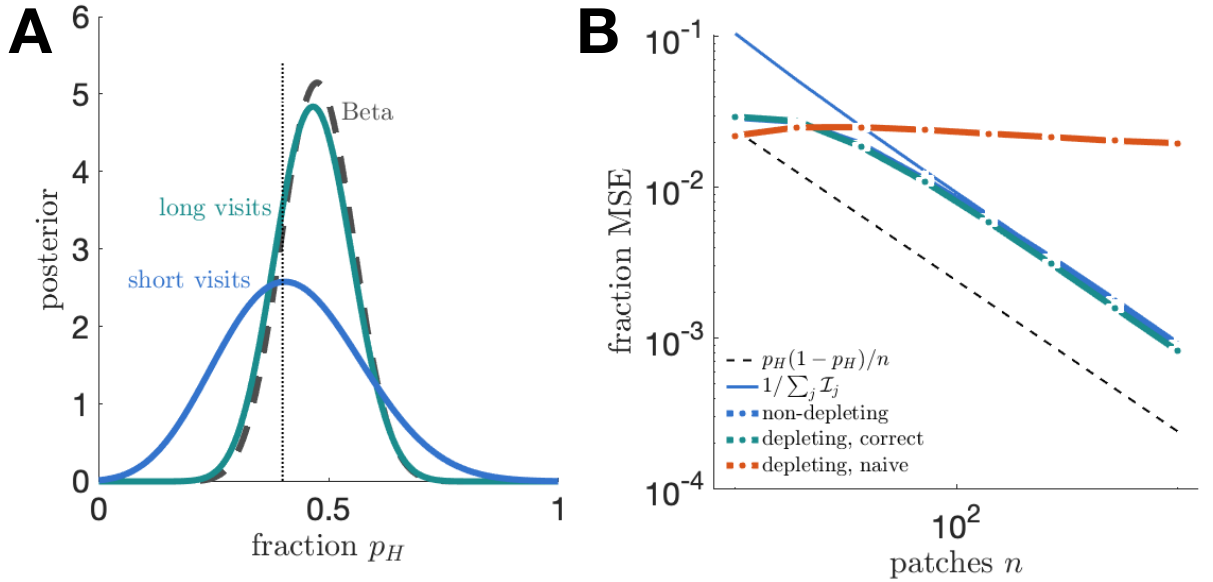} \end{center}
\vspace{-3mm}
\caption{\textbf{The patch-type fraction is learned slowly, and depletion does not affect it.} \textbf{A.} Posterior over the fraction $p_H$ with rates known, from one realization in which $19$ of $40$ visits were high. In the long-visit limit the exact soft-count posterior (teal) matches the conjugate Beta (dashed); short visits leave patch types ambiguous and broaden it (blue). True $p_H$ dotted. \textbf{B.} Fraction-estimate MSE versus number of patches $n$. Non-depleting (open markers) and depleting scored with the correct depleting likelihood (teal) lie on top of each other, decaying in parallel a factor of four above the rates-known $p_H(1-p_H)/n$ reference (dashed), the offset being the effective sample size $n_{\rm eff} < n$ of Eq.~(\ref{neff}) at this separation. The Fisher prediction $1/\sum_j {\mc I}_j$ (line) is asymptotic and exceeds both at small $n$, where a flat prior over $p_H$ holds the posterior mean near its center (S2~Text). Depleting counts scored with the naive non-depleting Poisson likelihood (red) instead plateau. Here $p_H = 0.4$.}
\label{fig3_composition}
\end{figure}

{\em Clean limit.} When a visit is long enough, or the rates well enough separated, that its type is resolved with near certainty, each patch is a Bernoulli trial with success probability $p_H$. With a conjugate Beta prior $p_H \sim {\rm Beta}(a_0, b_0)$, after $n$ visits of which $h$ are high the posterior is
\begin{align}
    p(p_H \mid h, n) = {\rm Beta}(a_0 + h, \, b_0 + n - h), \label{betapost}
\end{align}
with mean $\hat{p}_H = (a_0 + h)/(a_0 + b_0 + n)$. Taking $h \sim {\rm Binomial}(n, p_H^{\rm true})$ and a flat prior, the expected squared error decays as
\begin{align}
    \langle (\hat{p}_H - p_H^{\rm true})^2 \rangle = \frac{p_H^{\rm true}(1-p_H^{\rm true})}{n} + O(n^{-2}), \label{phmse}
\end{align}
the $1/n$ decay of a Bernoulli-rate estimate. This is the central contrast, since the rate MSE falls fast within a single patch, driven by the falling rate at every chunk, while the fraction MSE falls only as $1/n$ in {\em patches}.

{\em Noisy limit.} When visits are short or the rates close, a single visit does not resolve its type. Visit $j$ instead yields a soft type-posterior from its within-patch LLR $y_j = K_j \log(\lambda_H/\lambda_L) - (\lambda_H - \lambda_L) t_j$, giving high-type weight $\sigma_j = (1+e^{-y_j})^{-1}$, and the fraction posterior becomes an exact product of per-patch mixtures,
\begin{align}
    p(p_H \mid x_{1:n}) \propto p_0(p_H) \prod_{j=1}^n \left[ p_H\, \ell_H^j + (1-p_H)\, \ell_L^j \right], \label{softpost}
\end{align}
where $\ell_k^j = p(x_j \mid \lambda_k)$ is the likelihood of visit $j$'s data under type $k$. Conjugacy breaks, since each factor is linear in $p_H$ rather than a clean $p_H$ or $(1-p_H)$ (Methods). A resolved patch carries the full information of a clean Bernoulli trial while an ambiguous one carries almost nothing, so the per-patch Fisher information sums to an effective sample size
\begin{align}
    n_{\rm eff} = \sum_{j=1}^n (2\sigma_j - 1)^2 \leq n, \label{neff}
\end{align}
equal to $n$ only when every type is certain. Type confusion thus compounds the already-slow $1/n$ learning, and Eq.~(\ref{betapost}) is recovered as the rates separate and $n_{\rm eff} \to n$ (Fig.~\ref{fig3_composition}A). At the separation of Fig.~\ref{fig3_composition}B the shortfall is about fourfold.

{\em Depletion is neutral for fraction learning.} Since $n_{\rm eff}$ turns on how cleanly each visit reads its type, one might expect depletion to slow composition learning by capping the encounter count. It does not, provided the forager judges each visit by Eq.~(\ref{homlearndepj}), because the count is itself diagnostic of type, a high patch beginning with more chunks than a low one, so a visit long enough to deplete the patch reads its type at least as cleanly as a non-depleting visit of the same duration. Such a forager estimates the fraction without bias and its MSE lies on top of the non-depleting curve (Fig.~\ref{fig3_composition}B).

The neutrality is due to the mechanics of the inference model rather than the environment, which matters for how such data are analyzed. Fitting a non-depleting Poisson likelihood to depleting counts is misspecified, since depletion yields systematically fewer encounters than that model expects, so every patch reads as poorer than it is, high patches are confused with low ones, and the resulting fraction estimate is biased by an amount that does not shrink with more patches (Fig.~\ref{fig3_composition}B). An analysis of this kind would report that depletion degrades composition learning, an artifact of the likelihood rather than a property of the environment. Joint inference of the rates and the composition is treated in S2~Text.

\subsection{Reward versus information seeking}
\label{sec:rewardinfo}

Learning is coupled to movement, since the agent's departures generate the encounter stream it learns from. We now ask how a forager maximizing energy intake departs differently from one seeking to resolve the environment, first in the homogeneous environment where the uncertain quantity is the rate, then in the binary environment where the rates are known and only the composition remains.

\begin{figure}[t!]
\begin{center} \includegraphics[width=\textwidth]{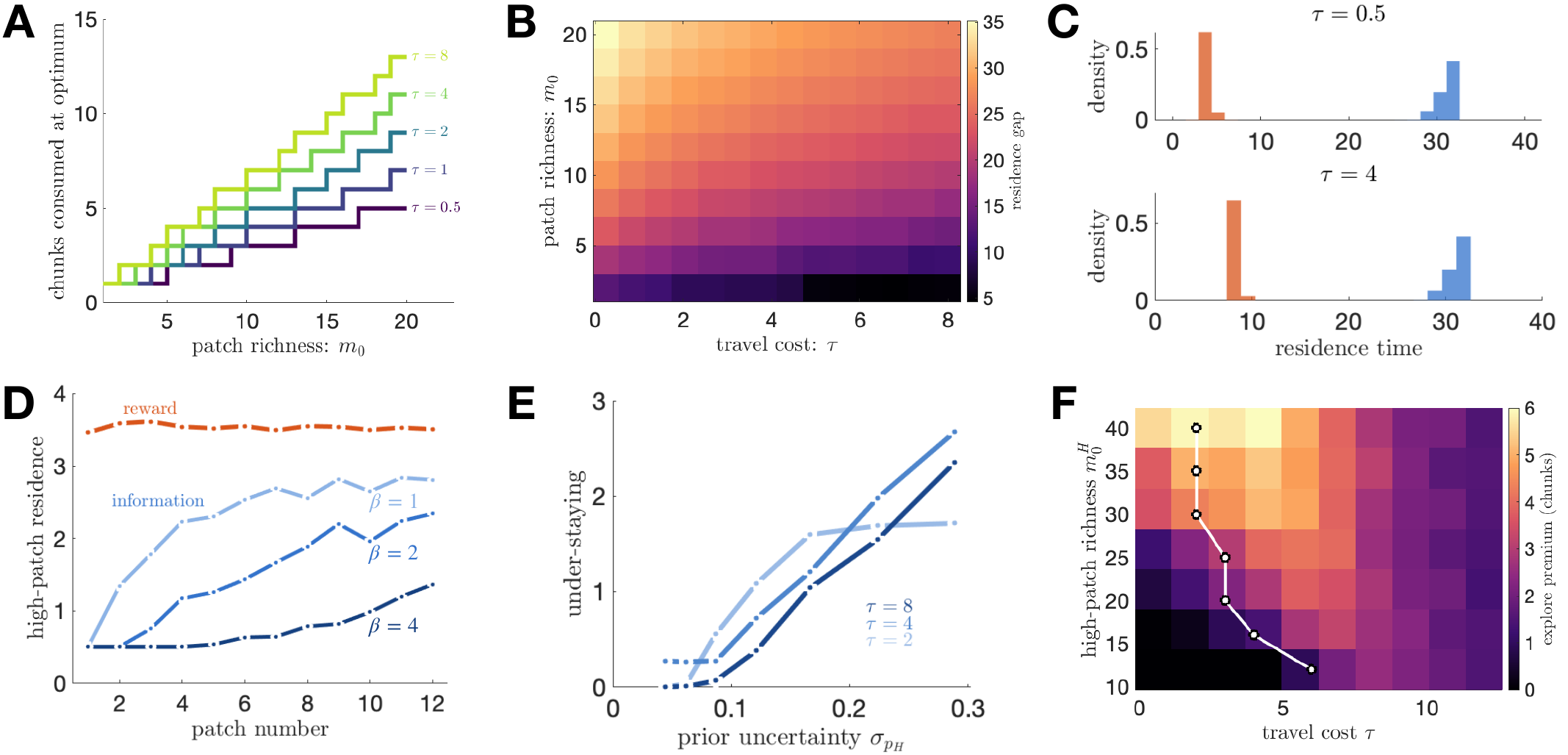} \end{center}
\vspace{-3mm}
\caption{\textbf{Reward- and information-seeking foragers diverge in observable patch-leaving behavior, and the direction of the divergence depends on what remains uncertain.} \textbf{A--C, homogeneous environment, rate uncertain.} \textbf{A.}~Chunks consumed at the reward optimum versus richness across travel costs, approaching the information-seeker's near-complete consumption at large $\tau$. \textbf{B.}~Residence-time gap between information- and reward-seekers over the travel cost and richness plane, set mainly by richness, largest for rich patches and cheap travel, and positive everywhere. \textbf{C.}~Residence-time distributions at $\tau=0.5$ (top) and $4$ (bottom), reward-seeker red and information-seeker blue. The two are disjoint at both travel costs. The information-seeker's rule does not reference $\tau$, so it consumes the patch and departs near $31$ in each, while the reward-seeker departs at $3.5$ and $8.2$ respectively. \textbf{D--F, binary environment, rates known and composition uncertain.} \textbf{D.}~High-patch residence by bout position at information weights $\beta = 1$, $2$ and $4$ (light to dark). The reward-seeker is flat, varying by under five percent of its mean across the bout, while every information-seeker begins pinned at the same departure floor and climbs toward it as $p_H$ resolves, the more heavily weighted more slowly. \textbf{E.}~The under-staying gap versus prior uncertainty $\sigma_{p_H}$ at three travel costs, every curve falling toward zero as the prior tightens. \textbf{F.}~The explore premium, chunks the reward-seeker consumes beyond the information-seeker, over the travel cost and high-patch richness plane, with the per-richness peak traced. The premium is largest at intermediate travel cost, and the ridge grows and moves toward cheaper travel as the rate gap widens, from $\tau=6$ at $m_0^H=12$ to $\tau \approx 1$ at $m_0^H=40$, where the premium is flat enough across $\tau$ that the traced peak is only weakly determined.}
\label{fig4_rewardinfo}
\end{figure}

\subsubsection{Homogeneous environments: the rate is uncertain}

We recall the reward-maximizing strategy derived previously for a single patch type~\cite{kilpatrick2021uncertainty,kilpatrick2020normative}. If the chunks remaining $m(t) = m_0 - K(t)$ were known exactly, a departure threshold $m_\theta$ on that remaining count, complementary to the consumed count $K_\theta$ of Section~\ref{sec:rates}, would generate
\begin{align}
    R^{m_{\theta}}(m_0) = \frac{m_0 - m_{\theta}}{\rho^{-1} \cdot \left[ H_{m_0} - H_{m_{\theta}} \right] + \tau}, \label{homknownrr}
\end{align}
where $H_m$ is the $m$th harmonic number, and the reward-optimal threshold is $m_{\theta}^{\rm opt}(m_0) = {\rm argmax}_{m_{\theta}} R^{m_{\theta}}(m_0)$. Given uncertainty about $m_0$ described by $p(m_0|x(t))$, the agent computes the rounded mean estimate $\bar{m}_0$ and departs when $K(t) > \bar{m}_0 - m_\theta^{\rm opt}(\bar{m}_0)$. The information-seeker instead departs once the posterior standard deviation over $m_0$ falls below a fixed threshold, the accuracy criterion of Section~\ref{sec:rates} applied as a stopping rule rather than a fixed chunk count.

A tractable case makes the contrast explicit. With patches holding either one or two chunks under a flat prior and a fixed bout of length $T$, the posterior is binary, the LLR is $y(t) = K(t)\log 2 - \rho t$, and each strategy reduces to a give-up time. Comparing expected reward against expected posterior entropy (Methods) gives a clean ordering, $t_g^{\rm info} \geq t_g^{\rm rew}$ for all $\tau$, since once depletion has set in the marginal informational value of waiting decays more slowly than the marginal reward value. The gap is widest as $\tau \to 0$, when a fresh patch is nearly free, and closes as travel becomes costly.

The same ordering holds for richer patches. With $m_0 \in \{1,\dots,m_{\max}\}$ under a flat prior, the posterior over $m_0$ given $K(t)$ encounters is $p(m_0|K(t)) \propto [m_0!/(m_0-K(t))!]e^{-m_0\rho t}$ for $K(t) \leq m_0$, and the reward-optimal threshold trades the reward from additional chunks against the depletion-driven decline in encounter rate. The chunks consumed at that optimum grow with both richness and travel cost (Fig.~\ref{fig4_rewardinfo}A), approaching the information-seeker's near-complete consumption as travel becomes costly. Two observables quantify the divergence without access to the forager's beliefs, the chunks consumed before departure and the residence time. The residence gap is set mainly by richness, largest for rich patches and cheap travel and smallest, though still positive, in the poorest (Fig.~\ref{fig4_rewardinfo}B). The reward-seeker's median residence more than doubles as travel becomes costly, while the information-seeker's rule makes no reference to $\tau$, so its median holds just past the $H_{m_0}/\rho = 29.3$ needed to consume all $m_0 = 10$ chunks and the two distributions are disjoint at both travel costs (Fig.~\ref{fig4_rewardinfo}C). That is the policy the MSE-threshold measure of Section~\ref{sec:rates} prescribes, reached here from an independent criterion. Both observables are measurable in standard assays, and the second separates the two foragers without error.

\subsubsection{Binary environments: the composition is uncertain}

In the binary environment we take the rates as already resolved and ask how the two agents depart when only $p_H$ remains uncertain. This isolates a structural form of the explore/exploit tradeoff, since with the rates known, exploration cannot target option value and instead targets the generative structure of the environment itself. Writing the reward rate over the type mixture with departure thresholds on the two types,
\begin{align}
    R(m_\theta^H,m_\theta^L) = \frac{p_H \rho (m_0^H - m_\theta^H) + p_L \rho (m_0^L - m_\theta^L)}{p_H [H_{m_0^H} - H_{m_\theta^H}] + p_L [H_{m_0^L} - H_{m_\theta^L}] + \rho \tau}, \label{binrewrate}
\end{align}
the reward-optimal thresholds maximize Eq.~(\ref{binrewrate}) and depend on $p_H$ only through the mixture weighting of the two types, not through any drive to learn it.

An information-seeker that additionally values resolving $p_H$ departs rich patches {\em earlier} than the reward-seeker, reversing the homogeneous ordering. The reason is asymmetric in how the two uncertainties are reduced. Rates are resolved by dwelling, since within-patch encounters sharpen the rate posterior, but the fraction is resolved only by sampling more patches, since each visit contributes one categorical observation of its type regardless of how long the agent stays. Composition information is gathered by leaving, not by staying. We model this by crediting each departure with the expected entropy reduction in the Beta posterior over $p_H$ from one additional type observation,
\begin{align}
    \Delta H(a,b) = H[{\rm Beta}(a,b)] - \left[ \tfrac{a}{a+b} H[{\rm Beta}(a{+}1,b)] + \tfrac{b}{a+b} H[{\rm Beta}(a,b{+}1)] \right], \label{compinfogain}
\end{align}
and adding it to the marginal-value opportunity cost that triggers departure. In a patch holding $m(t)$ chunks the instantaneous intake rate is $\rho\,m(t)$, and the reward-seeker departs where that rate falls to the long-run average under its current belief $\hat{p}_H = a_n/(a_n+b_n)$. Departure also buys one further type observation, worth $\Delta H(a_n,b_n)$, but only after the travel delay is paid, so we credit it at rate $\Delta H(a_n,b_n)/(\tau+T_0)$ and weight it by $\beta$, giving
\begin{align}
    \rho\, m(t) = \frac{\hat{p}_H m_0^H + (1-\hat{p}_H) m_0^L}{T_{\rm ref} + \tau} + \frac{\beta\, \Delta H(a_n,b_n)}{\rho\,(\tau + T_0)}, \label{infodepart}
\end{align}
where $T_{\rm ref}$ and $T_0$ are fixed reference times and the departure count is clamped to $[1,m_0-1]$ (Methods). Positive $\beta$ moves departure earlier, and the reward-seeker is the $\beta=0$ case.

This produces a clean behavioral signature (Fig.~\ref{fig4_rewardinfo}D). The reward-seeker's high-patch residence is flat across the bout, varying by under five percent of its mean, since its opportunity cost stops moving once its belief about $p_H$ is centered. The information-seeker under-stays high patches early, when $p_H$ is most uncertain, and climbs toward the reward-seeker without reaching it within a twelve-patch bout at every information weight $\beta$ we tested. That the gap is an information drive rather than a fixed bias follows from its dependence on resolvable uncertainty (Fig.~\ref{fig4_rewardinfo}E), since tightening the prior over $p_H$ shrinks it toward zero, an agent that already knows $p_H$ having nothing to learn by leaving. The explore premium then peaks at intermediate travel cost (Fig.~\ref{fig4_rewardinfo}F). Nearly free travel has both agents leaving readily, very costly travel has both tolerating a long stay, and only in between does the reward-seeker stay while the information-seeker still departs to sample composition. The ridge moves toward cheaper travel as the high patch grows richer, which follows from where the departure clamp releases rather than from the information term itself (Methods).

\subsection{Replenishing environments}
\label{sec:replenish}

The environments so far have been memoryless in space, since every departure leads to a fresh patch and a visited patch is never seen again. Many foragers instead exploit a fixed set of patches at known locations that recover between visits, so revisiting a depleted patch becomes a viable alternative to seeking a new one~\cite{ollason1987learning,van2010state,zeraati2025optimal}. An agent tracking each patch's recovery state alongside the composition can then cycle back to rich patches as they replenish. We show that the reward-maximizing policy collapses onto a stable limit cycle whose period is set by how many high patches the agent holds rather than by the recovery rate, and that replenishment then splits the inference hierarchy as depletion did, accelerating rate learning and suppressing composition learning through one parameter.

\begin{figure}[t!]
\begin{center} \includegraphics[width=19cm]{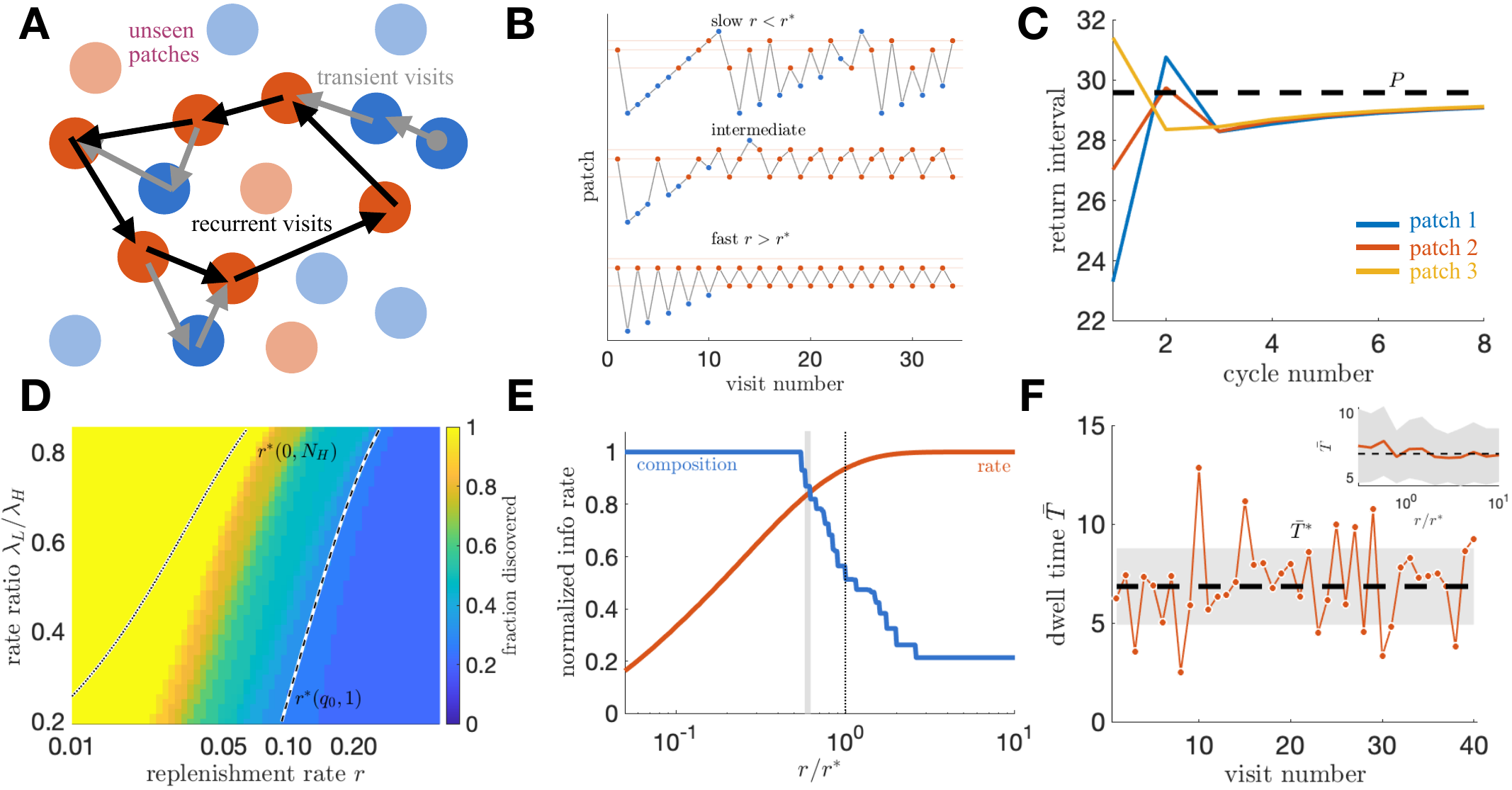} \end{center}
\vspace{-3mm}
\caption{\textbf{Replenishment sets the reward-seeker's orbit, the orbits attract, and replenishment splits rate learning from composition learning.} \textbf{A.}~Schematic of the arena and the orbit. Filled circles are patches the forager has typed, high red and low blue; pale circles are never reached, the pale red ones high patches it never finds, which set the floor on its composition estimate. Black arrows are the recurrent cycle, grey the transient visits preceding it. Positions are arbitrary, since travel cost is the same between every pair. \textbf{B.}~Three sample visit sequences in one arena at increasing replenishment rate, high patches red and low blue, connected in visit order, at a rate gap $\lambda_L/\lambda_H = 0.75$. Slow recovery drags the agent across the whole arena, intermediate gives a rotation through the high patches, fast collapses it onto two. \textbf{C.}~Per-patch return intervals from a schedule clumped into the first fifth of a period, relaxing onto the common period $P = N_H(\bar{T}^*+\tau)$ (dashed) to within $0.6$ percent of $P$ by the third cycle. \textbf{D.}~Fraction of patches discovered over the replenishment rate and the rate ratio, with the confinement band of Eq.~(\ref{rstar}) bracketed by a full rotation at zero belief, $r^*(0,N_H)$ (dotted), and a single recovered patch at the prior belief $\hat{q}_0 = 0.5$, $r^*(\hat{q}_0,1)$ (dashed). \textbf{E.}~The split. Fisher information per unit time about the cap $\lambda_H^{\max}$, learned on return, rises with $r$ (red), while information about the composition, gathered only at untyped patches, falls (blue). Both normalized to their maxima, $r = r^*$ dotted, the discovery step containing the crossing shaded. \textbf{F.}~Stochastic orbit at $r = 3r^*$. The departure condition becomes a threshold on a discrete chunk count, so the dwell time is a sum of independent exponentials with mean $4.2$ percent below the continuous $\bar{T}^*$, within one standard error of the $2.4$ percent the discrete count predicts, and with no drift over $200$ visits ($0.95$ SE); the first $40$ shown, band one standard deviation. Inset, mean dwell time and standard deviation over a thirty-fold range of $r$, remaining within $15$ percent of $\bar{T}^*$ throughout.}
\label{fig5_replenish}
\end{figure}

We fix $N$ patches at known locations, $N_H$ of them high and $N_H$ unknown to the agent (Fig.~\ref{fig5_replenish}A). Each patch depletes within a visit and recovers between visits by exponential relaxation toward its cap,
\begin{align}
    \lambda_k(s) = \lambda_k^{\max} - \left( \lambda_k^{\max} - \lambda_k^{\rm dep} \right) e^{- r s}, \label{replenish}
\end{align}
where $s$ is the time since the agent last left patch $k$, $\lambda_k^{\rm dep}$ its rate at that departure, and $r$ is shared across patches. The agent knows the locations, tracks the time since each visit, maintains a posterior over $N_H$, and knows $r$. We work in the deterministic limit, in which the within-patch rate decays as $\lambda_k^{\rm in} e^{-\rho t}$ and type is read exactly on entry, so composition is the only uncertain quantity; noise is restored at the end of the section. The route changes qualitatively as recovery quickens (Fig.~\ref{fig5_replenish}B), from a sweep across the arena to a rotation through the high patches to a collapse onto two.

{\em The reward orbit.} A reward-maximizing agent returns to the most-recovered high patch it has identified. With identical high patches and equal travel $\tau$, symmetry makes the orbit a uniform cycle in which each of the $n_c$ patches it holds is revisited every period $P = n_c(\bar{T}+\tau)$, optimal among rotations. The marginal value theorem sets the dwell time, and the recovered initial rate cancels from that condition, leaving
\begin{align}
    \rho\,(\bar{T}+\tau) = e^{\rho\bar{T}} - 1, \label{dwelleq}
\end{align}
so within-patch behavior is universal, independent of the cap, the replenishment rate and the composition. Since Eq.~(\ref{dwelleq}) fixes $\bar{T}^*$ independently of both $r$ and $n_c$, every rotation exists at every recovery rate, and the agent faces a coexisting family of orbits rather than a single one. Given $\bar{T}^*$, the recovered re-entry rate follows from Eq.~(\ref{replenish}),
\begin{align}
    \lambda^{\rm in} = \lambda_H^{\max}\,\frac{1 - e^{-rs}}{1 - e^{-(\rho\bar{T}^*+rs)}}, \qquad R^* = \lambda^{\rm in}e^{-\rho\bar{T}^*}, \label{reentry}
\end{align}
rising toward the cap as $r$ grows. Each orbit is locally stable and anisotropically so, the recovery coordinate contracting within a single visit while phase takes tens of cycles to re-equalize an uneven schedule (Fig.~\ref{fig5_replenish}C, Methods and S3~Text).

{\em Confinement is a band, not a threshold.} The decision governing discovery is whether to return to a known high patch or venture to an unvisited one. A high patch revisited after a gap $s$ offers $\lambda_H^{\max}(1-e^{-rs})$, while an unvisited patch offers its full undepleted rate, expected under belief $q$ to be $\mathbb{E}[\lambda_{\rm unv}] = q\lambda_H^{\max} + (1-q)\lambda_L^{\max}$. An agent cycling through $n_c$ patches waits $s = n_c(\bar{T}^*+\tau)$ between returns, giving
\begin{align}
    r^*(q,n_c) = -\frac{1}{n_c(\bar{T}^*+\tau)} \ln\!\left( 1 - \frac{\mathbb{E}[\lambda_{\rm unv}]}{\lambda_H^{\max}} \right). \label{rstar}
\end{align}
Since every rotation already exists at every $r$, $r^*$ marks not the creation of a cycle but the point at which one becomes attracting against exploring. That point is not a single recovery rate, because Eq.~(\ref{rstar}) depends on the size $n_c$ of the rotation the agent currently holds and on its belief $q$ about what remains unvisited, both of which the bout itself sets. Since $r^*$ scales as $1/n_c$, a large rotation stops being worth sustaining at a lower recovery rate than a small one, so as $r$ rises the largest sustainable rotation steps down through the integers. Confinement therefore occupies an interval of recovery rates rather than a threshold, running from $r^*(0,N_H)$, where even a full rotation through every high patch is too slow to be worth returning to and the agent is dragged to every patch, up to $r^*(\hat{q}_0,1)$, where a single recovered patch already beats an unvisited one at the prior belief and exploration stops with the first high patch found. Between them the agent explores just long enough to assemble a rotation that sustains itself, so how much of the arena it ever sees depends on how many high patches it happened to find on the way, and both edges descend through the bout as belief falls with experience (Fig.~\ref{figS3_replenish}A). The numerical transition falls inside the band at all $60$ rate ratios tested.

Discovery is complete at slow recovery and the fraction of the environment discovered decreases toward a floor of $(N+1)/[(N_H+1)N]$ (Fig.~\ref{fig5_replenish}D) as the recovery rate is increased. No higher recovery rate drives it lower, because an agent that has not yet found a high patch has nothing worth returning to and must keep exploring until it does, so the floor is the fraction of the arena that first search costs on average. Within the band the reward agent sustains a high-patch cycle but never characterizes the low patches, and that sorting comes from value-sensitive return rather than the recovery dynamics, since recovery fraction is type-independent and an agent returning to whichever patch has recovered most visits high patches only at their base rate $N_H/N$. Confinement is therefore a property of reward-maximizing return, not of the environment. An information-seeker weighs untyped patches against typed ones rather than recovered rates against unvisited ones, so no recovery rate confines it and its coverage stays complete at every $r$ (S3~Text). Raising the recovery rate should thus shrink a reward-seeker's range of exploration while leaving an information-seeker's behavior unchanged, a divergence a coverage measure would detect directly.

{\em Confinement is metastable once encounters are stochastic.} The analysis above compares deterministic values, so the orbit an agent settles into is fixed once its map is. Drawing the dwell time from the encounter process instead raises the discovered fraction in every combination of replenishment rate and rate ratio tested, by between $0.02$ and $0.19$, and lowers the squared error of the posterior mean composition in ten of twelve, since variance in the dwell time desynchronizes the return ages, breaks the rotation and pushes the agent back out to untyped patches. The deterministic completeness is therefore a lower bound, and $r^*$ separates regimes of escape rate rather than marking an absorbing boundary. Two noise sources act oppositely, dwell time variance aiding discovery while misreading a type on entry impedes it (S3~Text).

{\em Replenishment splits the hierarchy the way depletion does.} The mechanism is the same within-visit information, attenuated on re-entry rather than read at full strength on first arrival. A patch re-entered after a gap $s$ starts at the rate in Eq.~(\ref{reentry}), so an agent inferring the cap sees it only through a recovery-dependent sensitivity entering the Fisher information as its square (Methods). A barely recovered patch is nearly silent about its own cap, every encounter being drawn from a rate the recovery has scaled down, while a fully recovered one reports it directly, so rate information per unit time rises with $r$ and saturates once $rs \gg 1$. Composition information runs the other way, since each newly typed patch contributes one categorical observation as in Section~\ref{sec:composition} and the agent reaches untyped patches ever less often through the band, reaching zero once confined. Rate information rises with $r$ while composition information falls, so no recovery rate serves both levels for a forager maximizing intake, the two crossing within the discovery step spanning $0.62$ to $0.67\,r^*$ (Fig.~\ref{fig5_replenish}E).

Depletion and replenishment are thus complementary environmental mechanisms. Learning within a patch is accelerated by anything making the encounter stream more reliably report the patch's initial state and learning across patches by anything moving the agent to a patch it has not yet typed. Depletion supplies the first for free and is neutral for the second, since the interval spacing that pins the initial rate says nothing about how many patches share it. Replenishment supplies the first and destroys the second, so a reward-seeking forager in a fast-recovering environment ends up knowing the yield of a few patches precisely but does not learn the composition well. An information-seeker resolves $\hat{N}_H$ sooner below the band and permanently sooner above it, where the reward-seeker plateaus at the floor set by patches its cycle never reaches (S3~Text). None of this is an artifact of the deterministic limit, since restoring noise leaves the cycle intact, the dwell time falling $4.2$ percent below $\bar{T}^*$ without drift over $200$ visits and the prediction holding to within $15$ percent across a thirty-fold range of recovery rates (Fig.~\ref{fig5_replenish}F).

\subsection{Patch variability changes best strategy from counting to timing}
\label{sec:weakinstrument}

Everything so far has compared foragers by what they know. What that affords a forager in terms of intake depends on what they could have done instead, so we now compare against the best rule that never represents generative features of patches. The comparison yields a behavioral prediction, since the departure rule maximizing intake changes form with the variability of the patches, and a bound on what representing the environment is worth.

{\em One family contains every heuristic.} A departure rule answers one question, whether to stay given that $k$ chunks have been taken and none has arrived for $u$ time units, so any rule consulting nothing else is a single function $g(k)$, the foodless gap tolerated after the $k$th chunk. A fixed giving-up time is $g(k)$ constant, a fixed count of $m$ is $g(k)$ large below $m$ and zero at it, and the incremental and decremental rules of the classical literature are $g$ rising or falling in $k$~\cite{waage1979foraging,iwasa1981prey}. Because the process between chunks is Poisson at a known rate, expected intake under any $g$ follows in closed form, so we optimize over the function rather than argue over the names and the resulting bound is exact (Methods). Intake is reported throughout against an oracle that knows each patch's type on entry (Methods).

\begin{figure}[t!]
\begin{center} \includegraphics[width=19cm]{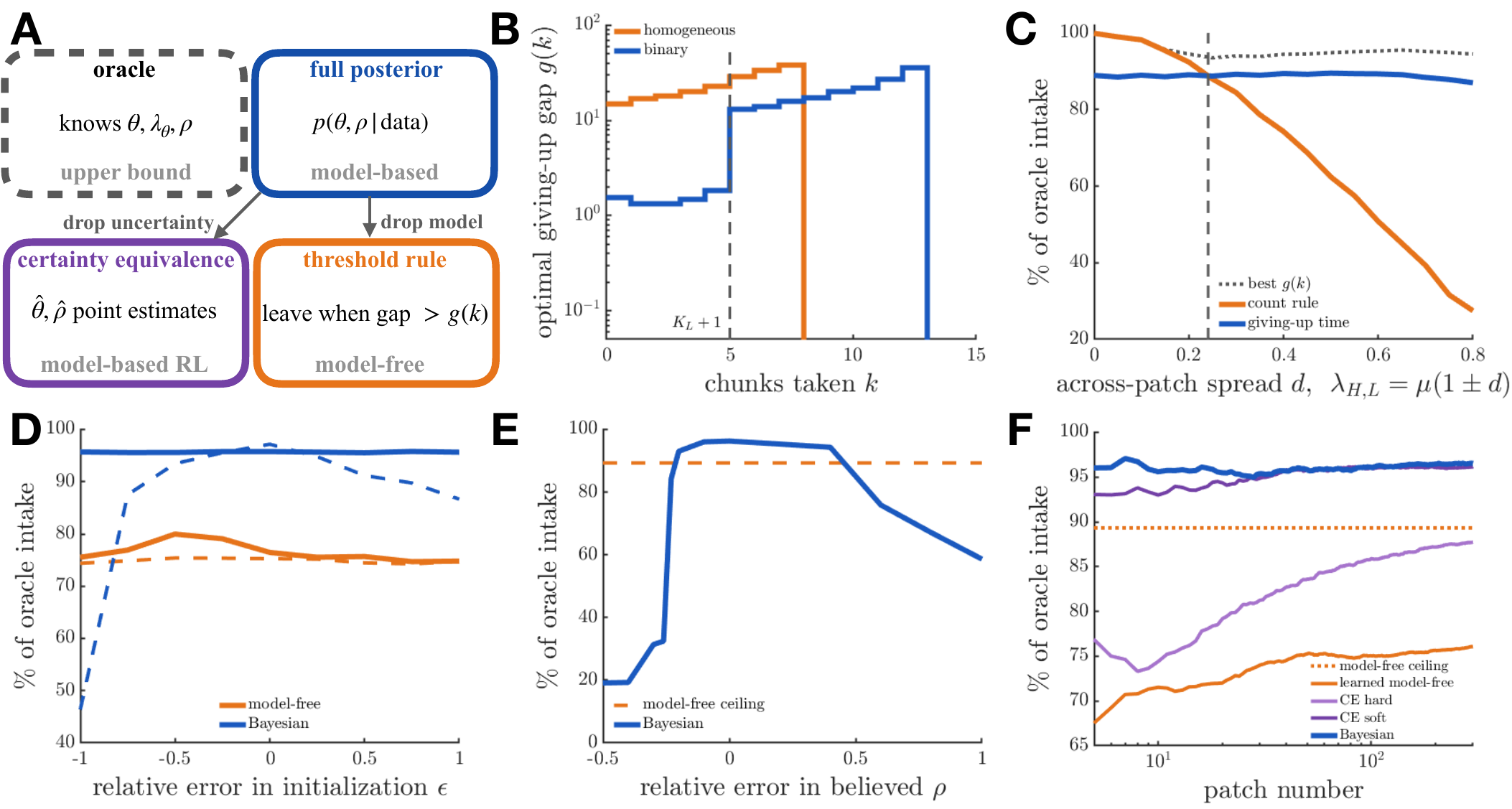} \end{center}
\vspace{-3mm}
\caption{\textbf{Variable patches favor timing over counting, and a model pays until it is wrong.} Binary depleting environment, $\lambda_H = 4$, $\lambda_L = 1$, $\rho = 0.25$, $\tau = 5$ unless noted; intake is reported against an oracle that knows the patch type. \textbf{A.} The four agent classes. Certainty equivalence is the posterior with uncertainty discarded, the threshold rule is the posterior with the model discarded, and the oracle is an upper bound rather than a parent. \textbf{B.} Unconstrained optimal giving-up gap $g(k)$, homogeneous (orange, $\lambda = 2.5$) and binary (blue), computed exactly, attaining $99.9$ and $95.3$ percent of the oracle respectively. Dashed line marks $m_0^L + 1$, one past the capacity of a poor patch. \textbf{C.} Best count rule and best fixed giving-up time against across-patch spread $d$, with $\lambda_{H,L} = \mu(1 \pm d)$ and $\mu = 2.5$, the unconstrained optimum dotted. The count rule falls from $99.9$ to $27$ percent across the range while the giving-up time holds near $89$ percent; dashed line marks the crossing at $d = 0.241$. \textbf{D.} Intake against matched relative error $\epsilon$ in each family's learned scalar, the opportunity cost for the posterior agent and the arm values for the model-free learner, converging to $95$--$96$ and $74$--$80$ percent respectively. Solid, converged window; dashed, first $50$ patches. $100$ realizations. \textbf{E.} Intake against relative error $m$ in the believed depletion rate, with prior support narrow enough to exclude the truth for $|m| > 0.15$. The band above the best gap-based rule ($89.3$ percent, dashed) runs from $m = -0.22$ to $+0.47$, peaking at $96.3$ percent. $50$ realizations. \textbf{F.} Learning curves at matched initialization for the full posterior, two certainty-equivalence variants, and the learned model-free agent, with the exact model-free ceiling dotted. $200$ realizations.}
\label{fig6_measure}
\end{figure}

{\em The optimum recovers the classical rules.} The best $g(k)$ takes a different shape in each environment (Fig.~\ref{fig6_measure}B). In the homogeneous case tolerance grows for moderate chunk counts and then collapses, recovering the marginal value count rule and nearly attaining the oracle. In the binary case tolerance is short and flat for small counts and then jumps by an order of magnitude, one chunk past the capacity $m_0^L$ of a poor patch. A rule that never represents patch type therefore approaches the oracle only by encoding the type-discrimination boundary as a kink in its own shape.

{\em The crossing is the prediction.} Writing $\lambda_{H,L} = \mu(1 \pm d)$, the count rule is optimal in a uniform arena and degrades steeply as richness spreads, while the giving-up time is insensitive to spread throughout, the two crossing at $d = 0.241$ (Fig.~\ref{fig6_measure}C). A capture should therefore shorten the remaining stay in uniform patches and lengthen it once richness varies by more than about a quarter, so a forager switching between decremental and incremental responsiveness may be tracking environmental variance rather than changing strategy. This is the sign prediction of Iwasa and colleagues obtained as a limit of the present framework~\cite{iwasa1981prey}, testable by manipulating variance at fixed mean richness.

{\em Learners fall short of both ceilings.} Perturbing each family's learned scalar by the same fractional error (Methods) leaves converged intake flat for both and about twenty percentage points apart (Fig.~\ref{fig6_measure}D), so the gap reflects what each agent represents rather than how either was initialized. The model-free shortfall is search rather than policy, since the best member of its own family sits well above where the learner ends up. The posterior agent's advantage also survives a change in the cost of acting, since freezing either heuristic at the threshold optimal beforehand costs it over a tenth of attainable intake in both directions and both environments, each threshold encoding the travel cost rather than the patches~\cite{kacelnik1992psychological,constantino2015learning}, while the posterior agent need only re-estimate its opportunity cost~\cite{daw2011model}.

{\em The advantage is bounded by model error.} We now give the agent a belief over the depletion rate on a narrow support centered on $\rho(1+m)$, so it infers the rate as before but cannot represent the truth once $|m|$ exceeds the half-width (Methods). This separates misspecification from uncertainty, since a wide prior centered on the wrong value is corrected by experience and a narrow one excluding the truth is not. Intake still beats every gap-based rule while the belief is between about a quarter below the true rate and about half above it, and falls short outside that band (Fig.~\ref{fig6_measure}E). The two failures differ in mechanism. Overestimating depletion makes the agent leave early and costs intake smoothly, while underestimating it pushes the required departure count past the patch capacity so the leaving condition is never met, giving a threshold near $\rho_{\rm bel} = (\lambda_H - R^\ast)/m_0^H$ and a discontinuous collapse. Outside the band a forager carrying no model at all does better.

{\em The cost of certainty falls on states, not parameters.} Holding the environment fixed and varying only the estimator locates that cost precisely (Fig.~\ref{fig6_measure}F, Methods). An agent replacing its belief over the depletion rate with a maximum likelihood value loses $0.3$ percentage points, indistinguishable from the full posterior, while one additionally committing to a patch type loses a further $8.8$ and falls below the model-free ceiling. The expensive approximation is not the point estimate of a parameter fixed across patches, but the point estimate of a state redrawn at every one, where a wrong commitment is paid immediately and cannot be averaged away. A model-based reinforcement learner therefore inherits the posterior agent's robustness to a change in travel cost, since both hold a model the manipulation does not touch~\cite{zeraati2025optimal}, but not its within-patch performance unless it also carries uncertainty over the current patch.

Three limits bound theses results. The discontinuous collapse under underestimated depletion is specific to the linear geometry, in which a patch holds $m_0^\theta = \lceil \lambda_\theta / \rho \rceil$ chunks and then yields nothing, though the tolerance band itself would survive a decline that never reaches zero. A more elaborate model-free agent, interpolating across travel costs it has not experienced or detecting a change and transiently raising its learning rate, would need to know how the environment generates reward, which is carrying a model in a different guise~\cite{nassar2010approximately}. And the comparison is at $N_T = 2$ in depleting environments, with S4~Text carrying it to eleven patch types and to model classes coarser than the environment.

Table~\ref{tab:taxonomy} collects the results across all six settings. Which departure rule a forager should use is set by how variable the patches are, not by how rich they are, and representing the environment buys about twenty points of intake over an agent that must learn its rule, but only while the depletion rate is believed to within roughly a quarter below and a half above the truth. Two things follow. Reward does not distinguish representations at their ceilings, since the best gap-based rule and the posterior agent lie within half a percentage point of each other, yet it separates by twenty points the agents that must find them, so a flat reward surface is a statement about optima and not about foragers. And asking whether an animal is model-based matters less than asking how wrong its model can be before the question stops mattering.

\begin{table}[t!]
\centering
\small
\setlength{\tabcolsep}{4pt}
\begin{tabularx}{\textwidth}{@{}>{\raggedright\arraybackslash}p{2.7cm}LLL>{\raggedright\arraybackslash}p{2.6cm}@{}}
\toprule
{\bf Setting} & {\bf Rate learning} & {\bf Composition learning} & {\bf Behavioral signature} & {\bf Figure} \\
\midrule
Homogeneous, non-depleting & Slow; MSE decay linear in $\lambda_1^{\rm true}$; no gain from moving & Not applicable & No divergence, staying is optimal for both & Fig.~\ref{fig2_rates}A \\
\addlinespace
Homogeneous, depleting & Faster above $\lambda_1^{\rm true} \approx 0.5$; interval spacing pins $\lambda_1$; consume the patch & Not applicable & Information-seeker {\em over}-stays; gap set mainly by richness, and larger when travel is cheap & Figs.~\ref{fig2_rates}B--C, \ref{fig4_rewardinfo}A--C \\
\addlinespace
Binary, non-depleting & Occupied type resolves, under-sampled type does not & $n_{\rm eff} < n$ under type ambiguity; $1/n$ in patches only once types are certain & Departure is informative once the occupied type is resolved & Figs.~\ref{fig2_rates}D, \ref{fig3_composition} \\
\addlinespace
Binary, depleting & Faster still; low rates never fully excluded & Unchanged from non-depleting if scored correctly & Information-seeker {\em under}-stays rich patches to sample types & Figs.~\ref{fig2_rates}E, \ref{fig4_rewardinfo}D--F \\
\addlinespace
Replenishing & Rises with $r$; saturates once $rs \gg 1$ & Falls with $r$; zero once the reward-seeker is confined & Reward orbit is one of a coexisting family; $r$ does not affect coverage of the information-seeker & Fig.~\ref{fig5_replenish} \\
\addlinespace
Binary depleting, departure rules compared & Not the object of comparison; the rates are known & Not applicable & Counting wins in uniform arenas, timing once richness spreads; a model beats no model unless badly misspecified & Fig.~\ref{fig6_measure} \\
\bottomrule
\end{tabularx}
\caption{{\bf Summary of results across settings and levels of inference.} Rows are the settings treated in Sections~\ref{sec:rates}--\ref{sec:weakinstrument}. Rate learning is inference of the within-patch encounter rate $\lambda_1$, scored by the mean squared error (MSE) of the posterior mean against the true value $\lambda_1^{\rm true}$; composition learning is inference of the fraction of patch types from $n$ visits, of which $n_{\rm eff}$ are informative when the occupied type is uncertain. In replenishing environments $r$ is the replenishment rate and $s$ the time away from a patch. Depletion acts within patches and not across them, the direction of the reward-versus-information divergence follows whichever quantity is uncertain, and the intake advantage of representing the environment holds only while the model of it is close enough to right.}
\label{tab:taxonomy}
\end{table}

\section{Discussion}

Learning the structure of the environment is among the most fundamental tasks an animal or artificial agent faces on entering a novel setting. Our central finding is that depletion, ordinarily treated only as the cost of exploiting a patch, is also a source of information, since each encounter both spaces and re-spaces the next. Composition, the fraction of patches of each type, is instead resolved one categorical observation at a time however richly each visit is sampled, so depletion neither helps nor hurts it. Rate learning therefore depends on what the encounter stream reports about the patch at hand and composition learning on how often the agent moves to a patch it has not yet typed, and coupled to departure rules weighing intake against information gain, the two strategies diverge most strongly early in exposure.

A subtler form of this tension arises within information seeking itself, where two reasonable measures of learning prescribe opposite movement in the homogeneous environment. That two learning objectives can disagree on whether to stay or go cautions against treating information seeking as a single well-defined drive, and suggests that distinguishing learning strategies empirically may require resolving not just whether an animal seeks information but which statistic of its belief it acts to reduce. The disagreement disappears once patches differ in type, since the marginal uncertainty of an under-sampled rate cannot be reduced without visiting the type that bears it, and depletion sharpens this by resolving the occupied type quickly and shifting the binding uncertainty onto types not yet seen.

Estimation accuracy predicts reward intake poorly in several ways. Priors differing widely in the rate error they leave after a single patch cost almost nothing in long-run intake, since the same falling within-visit rate that accelerates learning also overwrites the prior within a few encounters, leaving early belief strongly prior-dependent and long-run intake almost prior-free (S1~Text). A forager estimating the rates of only the richest and poorest patch types likewise outgathers one carrying the correct model, since widely separated hypotheses resolve the occupied patch faster, provided the richness spread is narrow enough that misassigning the rest stays cheap (S4~Text). The reward surface is flat near its optimum, so precision beyond a coarse estimate buys little while conditions hold fixed. That flatness is a property of optimal policies rather than of foragers. Agents that must find those policies separate widely at a fixed travel cost, since a belief is updated faster than a threshold is searched for, and separate again when the cost moves, because the belief survives a manipulation that invalidates every learned threshold. A model therefore affords a forager both speed and knowledge transfer, and both against a bound, since a coarse model class wins only while the richness spread stays narrow and a badly misspecified depletion rate loses to a rule carrying no model at all.

Our learning model occupies a level above the inference problems classical foraging theory has addressed. The marginal value theorem and its stochastic generalizations prescribe optimal residence given known environmental statistics~\cite{charnov1976optimal,oaten1977optimal}, and our prior work relaxed this to inference of the current patch type under a known type distribution~\cite{kilpatrick2021uncertainty,davidson2019foraging}. Here the type distribution itself must be estimated from the encounter stream, so the agent infers hierarchically over three nested quantities, the current patch type, the rates and fractions defining the environment, and, through the choice of departure rule, what to optimize. Framing learning this way clarifies that depletion enters at the rate-estimation level rather than only as a reward cost, and much of the hierarchy stays analytically tractable, yielding closed forms for the mean learning curve and for gains and departures as first-passage quantities.

This connects our results to a broader effort to understand foraging under incomplete and changing knowledge. Recent work argues that overharvesting, long treated as a deviation from optimality, can instead reflect rational structure learning by a forager not assuming complete knowledge~\cite{harhen2023overharvesting,harhen2026structure}, and behavioral studies of patch leaving under meta-uncertainty find that animals track local variability and global statistics with a hierarchical estimator~\cite{webb2025foraging}. Earlier behavioral-ecology treatments of Bayesian foraging reached related conclusions in restricted settings, including the two-patch-type case closest to our binary environment~\cite{olsson2006bayesian,mcnamara2006bayes,biernaskie2009bumblebees}. Our contribution is an analytic account of why the early phase of such learning is fast under depletion, and of how the movement it prescribes depends on whether the agent reduces reward or environmental uncertainty. Our flat priors are a limiting case rather than a neutral one, since cognitive biases from movement and spatial memory act as informative priors over the resources available~\cite{beardsworth2021habitat,gil2009honeybees}. A natural next step is whether the overharvesting signature survives when the cost of staying is weighed against the information a depleting patch's slow tail still supplies about the shared rate.

Foraging is also widely framed as reinforcement learning, in which an agent learns a leaving policy from experienced rewards rather than maintaining a posterior over environmental parameters~\cite{Kolling_Akam_2017,sutton2018reinforcement,mobbs2018foraging,zeraati2025optimal}. Section~\ref{sec:weakinstrument} shows the framings are behaviorally separable, though not where the labels suggest. At their ceilings they are indistinguishable, the best rule available to an agent with no representation of the patches landing within half a percentage point of the Bayesian agent, while the agents that must find those rules differ by about twenty points. Nor does holding a model suffice on its own. An agent planning against point estimates matches the full posterior when the estimate is of a parameter fixed across patches, losing $0.3$ percentage points, and falls below even the model-free ceiling when it also commits to the current patch type, losing a further $8.8$. Model-based reinforcement learning therefore inherits the posterior agent's robustness to a change in travel cost, since both hold a model the manipulation does not touch~\cite{zeraati2025optimal}, without inheriting its within-patch performance. What matters is not whether an agent holds a model but whether it carries uncertainty over the quantity redrawn at every patch.

Several of these questions are measurable in experiments. Whether a forager reduces reward or environmental uncertainty should be visible in residence times and in sensitivity to travel cost, and our homogeneous result shows the two can diverge even when both are framed as information seeking, so an experiment varying depletion rate and travel time across early and late visits could test whether animals shift from information-led to reward-led departures as the environment resolves. Uncertain travel should widen the model-based advantage rather than blur it, since a transit cost redrawn at every departure adds its variance to the reward signal a model-free learner searches over while leaving a belief about the patches untouched, and the gap should widen further when the forthcoming transit is visible beforehand, the optimal rule being a function of the observed cost that an agent holding a model computes directly and one without must learn separately for each cost. Neither case is simulated here. The framework extends in directions we have not pursued, including learning the depletion increment through the same sequential updates, continuous spatial landscapes where exploration and route planning interact, and social settings in which an agent infers the parameters governing other foragers~\cite{giraldeau2018social,perez-escudero_collective_2011}.

\section*{Methods}

\subsection*{Reward rate over many visits}

The efficiency of a policy $\Pp$ is computed over the times spent $T_{1:n}$ and resources gained $r_{1:n}$ across $n$ patch visits as $R_n(\Pp) = \sum_{j=1}^n r_j/\left[ \sum_{j=1}^n [T_j + \tau_j] \right]$, where $\tau_{1:n}$ are the delays required to reach each patch from the previous. Scaling numerator and denominator by $n$ and taking the limit of many visits gives the means $R(\Pp) = \lim_{n \to \infty} R_n (\Pp) = \langle r \rangle_{\Pp} / [ \langle T \rangle_{\Pp} + \langle \tau \rangle_{\Pp}]$, which for a distribution $p(\lambda)$ is Eq.~(\ref{energygain}).

\subsection*{Rate estimation in non-depleting homogeneous environments}

From Eq.~(\ref{nondeplearn}), the MLE of $\lambda_1$ satisfies the implicit equation
\begin{align*}
    t \hat{\lambda}_1 &=  K(t) + \hat{\lambda}_1 p_0'(\hat{\lambda}_1)/p_0(\hat{\lambda}_1),
\end{align*}
which is consistent, since
\begin{align*}
    \lim_{t \to \infty} \hat{\lambda}_1 = \lim_{t \to \infty} \frac{K(t)}{t} +  \lim_{t \to \infty} \frac{\hat{\lambda}_1 p_0'(\hat{\lambda}_1)/p_0(\hat{\lambda}_1)}{t} = \lambda_1.
\end{align*}
For a flat improper prior $p_0(\lambda_1) \equiv 1$ this gives $\hat{\lambda}_1 = K(t)/t$, and for an exponential prior $p_0(\lambda_1) = t_0 e^{-t_0 \lambda_1}$ it gives $\hat{\lambda}_1 = K(t)/(t+t_0)$. The flat prior generates unbiased estimates since $E[\hat{\lambda}_1|\lambda_1] = \lambda_1 t/t = \lambda_1$, whereas the exponential prior generates a slight underestimate whose bias decays algebraically. The posterior under the exponential prior is Gamma$(K+1, t+t_0)$, with posterior mean
\begin{align*}
    \tilde{\lambda}_1 = \int_0^{\infty} \frac{(t+t_0)^{K+1}\lambda_1^{K+1}}{K!} e^{- \lambda_1 (t + t_0)} d \lambda_1 = \frac{K(t) + 1}{t + t_0},
\end{align*}
and hence bias
\begin{align*}
{\rm Bias} \left[ \tilde{\lambda}_1 | \lambda_1 \right] = \frac{\lambda_1 t + 1}{t + t_0} - \lambda_1 = \frac{1 - \lambda_1 t_0}{t + t_0}.
\end{align*}

\subsection*{Ensemble mean of the MSE}

To obtain Eq.~(\ref{msevthomno}) we average the posterior across realizations of the encounter process,
\begin{align*}
    \bar{p} (\lambda_1 | t) = E[p(\lambda_1 | x(t)) | \lambda_1^{\rm true}] &= \sum_{K=0}^{\infty} p(K(t)| \lambda_1) p(K(t) | \lambda_1^{\rm true}) \frac{p_0(\lambda_1)}{p(K)} \\
   &= \sum_{K=0}^{\infty} \frac{(\lambda_1^{\rm true} t)^K (\lambda_1 t)^K e^{- \lambda_1^{\rm true} t} e^{- \lambda_1 t} p_0(\lambda_1)}{(K!)^2 p(K)}.
\end{align*}
For a flat improper prior the marginal is
\begin{align*}
    p(K) = \int_0^{\infty} p(K | \lambda_1) d \lambda_1 =  \frac{t^K}{K!} \int_0^{\infty}  \lambda_1^K e^{- \lambda_1 t} d \lambda_1 = \frac{1}{t},
\end{align*}
so that $p(\lambda_1 | K(t)) = (t/K!) (\lambda_1 t)^{K} e^{- \lambda_1 t}$ and
\begin{align}
    \bar{p}(\lambda_1 | t) &= t e^{- (\lambda_1^{\rm true} + \lambda_1) t} \sum_{K=0}^{\infty} \frac{(\lambda_1^{\rm true} \lambda_1 t^2)^K }{(K!)^2} \nonumber \\
    &= t I_0(2t \sqrt{\lambda_1 \lambda_1^{\rm true}})  e^{- (\lambda_1^{\rm true} + \lambda_1) t} \equiv F(t,\lambda_1, \lambda_1^{\rm true}), \label{fpmeanmsehomno}
\end{align}
where $I_0(z)$ is the zeroth order modified Bessel function of the first kind. Integrating Eq.~(\ref{fpmeanmsehomno}) against the MSE formula and changing variables via $\lambda_1^{\rm true} x = \lambda_1$ and $s = \lambda_1^{\rm true} t$,
\begin{align*}
    \bar{\rm MSE}(t;\lambda_1) = \int_0^{\infty} \frac{(\lambda_1 - \lambda_1^{\rm true})^2}{(\lambda_1^{\rm true})^2}  F(t,\lambda_1, \lambda_1^{\rm true}) d \lambda_1 = \int_0^{\infty} (x - 1)^2  F(s,x, 1) d x,
\end{align*}
so the decay rate of the normalized mean MSE increases linearly in $\lambda_1^{\rm true}$.

\subsection*{Entropy and information gain in non-depleting environments}

For an exponential prior $p_0(\lambda_1) = t_0 e^{- \lambda_1 t_0}$, the posterior entropy after a time $t-t_0$ is
\begin{align*}
H(K, t, t_0) = - \int_0^{\infty} p(\lambda_1 | K(t)) \log p(\lambda_1 | K(t)) d \lambda_1 =  K+1 - \log \left[ t + t_0 \right] + \log K! - K \psi  \left( K+1 \right),
\end{align*}
where $\psi(z)$ is the digamma function. For $t \gg 1$ and $K \gg 1$, Stirling's approximation and an approximation of the digamma function give $H \approx - \log[t+t_0]$, decreasing indefinitely as the rate is learned. The information gained is the reduction in entropy,
\begin{align*}
    IG(K, t, t_0) = H(0, t_0, t_0)-H(K, t, t_0) = \log \frac{t+t_0}{t_0} + K \psi (K+1) - \log K! - K \to \log \frac{t+t_0}{t_0},
\end{align*}
so information gain scales logarithmically with time. Its rate has contributions both from chunk encounters at $t_k$ and from the time without food between them,
\begin{align*}
    \frac{d}{dt} IG(K, t, t_0) = \frac{1}{t+t_0} + \sum_{K=1}^{\infty} \left[ \log (K+1) - \psi (K+1) \right] \delta(t-t_k),
\end{align*}
which is exact, and which on averaging in the $t \gg 1$ limit gives
\begin{align*}
    \left \langle \frac{d}{dt} IG(K, t, t_0) \right \rangle \approx \frac{1}{t+t_0} + \frac{1}{2 t},
\end{align*}
independent of the true arrival rate. Higher rates deliver more rapid encounters, but precise identification of a higher rate is correspondingly harder.

\subsection*{Stay-versus-go information gain in binary environments}

Consider an agent that has visited only high patches and seeks the unobserved $\lambda_L$. Staying bears on $\lambda_L$ only through the residual probability that the occupied patch is low after all, and that probability falls exponentially as the visit lengthens, so the numerator of $IGR_{\lambda_L}^{\rm stay}$ in Eq.~(\ref{igrstay}) is $O(p_L e^{-(\lambda_H-\lambda_L)\Delta t})$, exponentially small in the rate gap. Departing instead carries probability $p_L$ of entering a low patch and resolving it, so the numerator of $IGR_{\lambda_L}^{\rm depart}$ is $O(p_L)$, bounded away from zero whenever the fresh patch can be the under-sampled type. The denominators differ only by the travel delay and the mean time to the first chunk in the fresh patch, so once the occupied type is resolved $IGR_{\lambda_L}^{\rm depart} > IGR_{\lambda_L}^{\rm stay}$. S4~Text plots the ratio against the rate gap.

\subsection*{Fraction posterior under type ambiguity}

Conjugacy breaks in Eq.~(\ref{softpost}) because each factor is linear in $p_H$ rather than a clean $p_H$ or $(1-p_H)$, so the posterior is a degree-$n$ polynomial in $p_H$, which we evaluate on a grid. The cost of type ambiguity is set by the per-patch Fisher information
\begin{align*}
    {\mc I}_j(p_H) = \frac{(\ell_H^j - \ell_L^j)^2}{\left( p_H \ell_H^j + (1-p_H)\ell_L^j \right)^2},
\end{align*}
whose reciprocal sum gives the asymptotic estimator variance. A resolved patch contributes the full $1/[p_H(1-p_H)]$ of a clean Bernoulli trial while an ambiguous one contributes almost nothing, so $p_H(1-p_H)\sum_j {\mc I}_j$ reads as the effective sample size of Eq.~(\ref{neff}).

\subsection*{Fisher information and the Jeffreys prior}

For a depleting Poisson process with initial rate $\lambda_1$ observed for a fixed time $t$, the probability of encountering $K$ chunks is
\begin{align}
    p(K(t) | \lambda_1) = \left( \begin{array}{c} \lambda_1/\rho \\ K(t) \end{array} \right) (e^{\rho t}-1)^{K(t)}e^{- \lambda_1 t}, \ \ \ \ K(t) \leq \lambda_1/ \rho.
\end{align}
This is the count-only likelihood, whereas the inference of Eq.~(\ref{homlearndepj}) uses the full arrival times, so the bound below is conservative relative to what the posterior achieves. The Jeffreys prior is the square root of the determinant of the Fisher information, so for this single-parameter problem $p(\lambda_1; \rho) \propto \sqrt{{\mc I}(\lambda_1; \rho)}$ with
\begin{align}
    {\mc I}(\lambda_1; \rho) &= \sum_{K=0}^{\lambda_1/\rho} \left[ \frac{d}{d \lambda_1} \log p(K| \lambda_1) \right]^2 p(K| \lambda_1) \nonumber \\
    &= e^{- \lambda_1 t} \sum_{K=0}^{\lambda_1/\rho} \left[ \sum_{j = \lambda_1/\rho -K+1}^{\lambda_1/\rho} \frac{1}{\rho j} - t \right]^2 \left( \begin{array}{c} \lambda_1/\rho \\ K \end{array} \right) (e^{\rho t} -1)^K. \label{depoisfi}
\end{align}
The encounter count has $\lambda_1$-dependent support, $K(t) \leq \lambda_1/\rho$, so we apply the Fisher information identity under the interior-support regularity condition, the boundary term from the moving upper limit vanishing since $p(K = \lambda_1/\rho \,|\, \lambda_1)$ enters the score symmetrically and cancels in the expectation. In the limit $\rho \to 0^+$ we recover the Fisher information of a non-depleting Poisson process,
\begin{align}
    \lim_{\rho \to 0} {\mc I}(\lambda_1; \rho) &= e^{- \lambda_1 t} \sum_{K=0}^{\infty} \lim_{\rho \to 0} \left[ \left( \frac{1}{\lambda_1} + \cdots + \frac{1}{\lambda_1 - \rho (K-1)} \right) - t \right]^2 \frac{\prod_{j=0}^{K-1} (\lambda_1 - \rho j)}{K!} \frac{(e^{\rho t}-1)^K}{\rho^K} \nonumber \\
    &= e^{- \lambda_1 t} \sum_{K=0}^{\infty} \left[ \frac{K}{\lambda_1} - t \right]^2 \frac{\lambda_1^K t^K}{K!} = \frac{t}{\lambda_1}, \label{poissfi}
\end{align}
using $\lim_{\rho \to 0} (e^{\rho t} -1)^K/\rho^K = t^K$. The Jeffreys prior of the homogeneous Poisson process is therefore invariant to measurement time, $p(\lambda_1; \rho =0) \propto \sqrt{1/\lambda_1}$, and improper since its integral diverges. S1~Text normalizes it over a finite interval to compare against alternatives.

\subsection*{The exclusion bound is not a separate source of information}

Each encounter in a depleting patch implies $\lambda_1 \geq K(t)\rho$, which truncates the likelihood of Eq.~(\ref{homlearndepj}). Ablating the two features separately shows the truncation contributes nothing on its own. Removing it while retaining the falling-rate form leaves the posterior error unchanged at $0.0036$, whereas retaining it while replacing the falling-rate form with a constant-rate one raises the error to $0.031$, an order of magnitude. What identifies $\lambda_1$ is the interval spacing, and the bound is a consequence of the same falling rate rather than an addition to it.

\subsection*{Reward and entropy for the two-chunk patch}

Take patches containing $m_0 \in \{1,2\}$ chunks with a flat prior, a bout of length $T$, and the LLR $y(t) = K(t)\log 2 - \rho t$. Under the strategy of staying until another chunk arrives ($t_g = T$),
\begin{align*}
    p(r_s = 2|t_g = T, t_1) &= \frac{2 e^{- \rho t_1}}{1 + 2 e^{- \rho t_1}} \left[ 1 - e^{- \rho (T - t_1)} \right], \qquad
    p(r_s = 1|\cdot) = 1 - p(r_s = 2|\cdot),
\end{align*}
since $r_s = 2$ requires both that the patch is type $m_0=2$, with posterior probability $2e^{-\rho t_1}/(1+2e^{-\rho t_1})$ after a first chunk at $t_1$, and that its second chunk, arriving at the depleted rate $\rho$, is encountered before the bout ends. The expected reward from staying is $\bar{r}_s = p(r_s=1) + 2p(r_s=2)$. Comparing against departing yields the rule that the forager waits for a second chunk only when $y(t_1) > 0$, equivalently $t_1 < (\log 2)/\rho$. Whether waiting improves expected reward depends on depletion rate and travel cost together, with staying favored once roughly $2\rho\tau > 1$~\cite{kilpatrick2020normative}.

For information seeking, the entropy over the LLR is
\begin{align}
    H(y) = \frac{\log (1 + e^{-y})}{1 + e^{-y}} + \frac{\log (1 + e^{y})}{1 + e^{y}} \equiv \log (1 + e^{-y}) + \frac{y}{1+e^{y}},  \label{entropy_LLR}
\end{align}
even in $y$ and decreasing as $|y|$ increases from zero, beginning at $H(0) = \log 2$. In the limit $t_g \to T$ the agent stays unless it encounters another chunk, which gives certainty $\lambda_1 = 2\rho$ with probability $(1 - e^{-\rho(T-t_1)})/(1+e^{-y_1})$, and otherwise leaves the entropy at $y = \log 2 - \rho T$. The expected entropy from staying is thus
\begin{align}
    \bar{H}_T^{\rm stay} = \frac{1 + 2 e^{- \rho T}}{1 + 2e^{-\rho t_1}} \cdot \left[ \log \left( 1 + \frac{1}{2} e^{\rho T} \right) + \frac{\log 2 - \rho T}{1 + 2 e^{- \rho T}} \right]. \label{entrop_stay}
\end{align}
For $t_g < T$ the agent may still encounter a chunk before $t_g$, with probability $(2e^{-\rho t_1} - 2e^{-\rho t_g})/(1+2e^{-\rho t_1})$; otherwise it travels and forages for $t_r = T - t_g - \tau$. Because the environment is homogeneous the second patch shares the type of the first, so it yields zero, one or two chunks with probabilities
\begin{align*}
    p(K_1(t_g) = 1, K_2(t_r) = 0 ) &= \frac{e^{-\rho t_r} + 2 e^{-\rho t_g} e^{-2\rho t_r}}{1 + 2 e^{-\rho t_1}}, \\
    p(K_1(t_g) = 1, K_2(t_r) = 1) &= \frac{\left(1 - e^{-\rho t_r}\right) + 4 e^{-\rho t_g}\left( e^{-\rho t_r} - e^{-2\rho t_r}\right)}{1 + 2 e^{-\rho t_1}}, \\
    p(K_1(t_g) = 1, K_2(t_r) = 2) &= \frac{2 e^{-\rho t_g}\left(1 - e^{-\rho t_r}\right)^2}{1 + 2 e^{-\rho t_1}},
\end{align*}
summing to $p(K_1(t_g)=1)$. The total count across both patches sets the final LLR $y = (1+K_2)\log 2 - \rho(T-\tau)$, so the expected entropy is
\begin{align}
    \bar{H}_T^{{\rm go}, t_g} = \sum_{k=0}^{2} p(K_1(t_g)=1, K_2(t_r)=k)\, H\!\left(y(K_2=k)\right), \label{entrop_go}
\end{align}
with $H(\cdot)$ from Eq.~(\ref{entropy_LLR}). Taking $t_g \to T$ in Eq.~(\ref{entrop_go}) and replacing $T-\tau$ with $T$, since the agent would then stay, recovers Eq.~(\ref{entrop_stay}). The reward-seeking give-up time maximizes $\bar{r}_s(t_g)$ and the information-seeking time minimizes $\bar{H}_T^{{\rm go},t_g}$.

\subsection*{Departure under a composition-information drive}

Eq.~(\ref{infodepart}) carries two fixed reference times. $T_{\rm ref}$ stands in for the reward-optimal residence in a high patch and enters only the opportunity cost, where it sets the scale of the long-run average. $T_0$ is a nominal within-patch dwell time that keeps the information credit finite as travel becomes free, since an agent able to reach a fresh patch at no cost would otherwise be credited without bound. Neither is fitted, and we hold $T_0 = 1$ in our time units throughout. Because $\Delta H$ is measured in nats while the remaining terms are chunks per unit time, $\beta$ carries units of chunks per nat per unit time.

Solving Eq.~(\ref{infodepart}) for the count at departure gives
\begin{align*}
    K_{\rm leave} = m_0 + 1 - \frac{1}{\rho}\left[ \frac{\hat{p}_H m_0^H + (1-\hat{p}_H)m_0^L}{T_{\rm ref}+\tau} \right] - \frac{\beta\,\Delta H(a_n,b_n)}{\rho^2 (\tau + T_0)},
\end{align*}
so the information term shifts departure by the same number of chunks in both patch types, and the explore premium is not itself richness dependent. The ridge of Fig.~\ref{fig4_rewardinfo}F comes from the clamp. When travel is cheap the opportunity cost is large and both agents sit at the floor, when it is costly both sit at the cap, and the premium is visible only in the window between, which opens toward cheaper travel as $m_0^H$ grows, since the floor releases once the opportunity cost falls below $\rho m_0$. The peak falls from $\tau = 6$ at the poorest high patch to $\tau \approx 1$ at the richest at every $T_0$ from $0.5$ to $8$, so the direction does not depend on the choice, though the traced peak shifts by up to one grid step across that range. That the mechanism is the clamp rather than the information term is visible at the poorest high patch, where the premium is identically zero for $\tau \leq 2$ because the opportunity cost holds both agents at the floor.

\subsection*{Rate information on re-entry in replenishing environments}
A patch re-entered after a gap $s$ is observed only through $\lambda^{\rm in}$, so an agent inferring the cap sees it through
\begin{align}
    \frac{\partial \lambda^{\rm in}}{\partial \lambda^{\max}} = \frac{1-e^{-rs}}{1-e^{-(\rho \bar{T}^*+rs)}}, \label{recovsens}
\end{align}
and the Fisher information such a visit carries about $\lambda^{\max}$ is this sensitivity squared times the single-visit information of Eq.~(\ref{depoisfi}) evaluated at $\lambda^{\rm in}$. The sensitivity rises from zero toward one as $rs$ grows, which is the quantity normalized in Fig.~\ref{fig5_replenish}E.

\subsection*{Stability of the return cycle}
Orbit stability is anisotropic. Because the departure rate equals the long-run cycle average whatever rate the agent arrived at, the recovery coordinate resets within a single visit, so a forager displaced from the patch set resumes its rotation in one cycle however saturated the patches have become. Phase is far weaker. A Floquet probe of the return map gives a leading multiplier of $0.98$ at $r = r^*$ with the opportunity cost held at its fixed point, so an unevenly spaced schedule takes tens of cycles to re-equalize. The faster relaxation in Fig.~\ref{fig5_replenish}C reflects an agent that also updates its opportunity cost from its own intake history, a mechanism the model as described does not include, so that panel demonstrates re-equilibration of the schedule rather than measuring phase stability.

\subsection*{Simulation checks on the replenishing dynamics}
The closed-form completeness underlying Fig.~\ref{fig5_replenish}D is checked against direct simulation of the same choice rule, which reproduces it to within $0.02$ in almost every combination of replenishment rate and rate ratio tested. Where noise is restored, the dwell time is drawn from the encounter process at the patch's actual recovered rate, the agent leaving at the residual $R^*$ of Eq.~(\ref{reentry}), so departure is a threshold on a discrete chunk count rather than a continuous time. Discovery comparisons average $60$ realizations at each of twelve combinations of replenishment rate and rate ratio, with the choice rule held identical to the deterministic case so that dwell time variance is the only difference.

\subsection*{Departure rules and agent classes}

\paragraph{Environment.} A patch of type $\theta$ delivers chunks as a Poisson process whose rate after $k$ chunks is $\lambda_\theta(k) = \max(\lambda_\theta - k\rho, 0)$, so the patch holds $m_0^{\theta} = \lceil \lambda_\theta/\rho \rceil$ chunks. Types are drawn independently across patches with probabilities $p_\theta$, and travel between patches costs $\tau$. Performance is long-run intake, total chunks divided by total elapsed time including travel. Departure is evaluated continuously, since a gap without food is itself evidence.

\paragraph{Oracle.} An agent knowing $\theta$, $\lambda_\theta$ and $\rho$ leaves when $\lambda_\theta(k)$ falls below the long-run intake rate $R^\ast$, which is fixed by the self-consistency condition of the marginal value theorem and solved by bisection. This is the upper bound against which every intake is reported.

\paragraph{Threshold family.} A rule that consults only $k$ and the elapsed gap $u$ is written $g(k)$, with departure at the first $u > g(k)$. A fixed giving-up time is $g$ constant, a fixed count of $m$ is $g$ large below $m$ and zero at it, and incremental and decremental rules are $g$ increasing or decreasing. Within a patch the process is Markov in $k$, so with $p_k = 1 - e^{-\lambda_\theta(k) g(k)}$ the probability of reaching stage $k+1$ and $\mathbb{E}[\min(X, g(k))] = p_k/\lambda_\theta(k)$ the expected time in stage $k$, expected chunks and expected residence are sums over stages weighted by $\prod_{i<k} p_i$. Intake follows exactly with no simulation, so the count and giving-up optima are found by one-dimensional search and the unconstrained optimum $g(k)$ by coordinate ascent over a grid spanning $0.02$ to $80$, iterated to convergence and truncated at the first count where the optimal tolerance collapses to a small fraction of the largest value reached, beyond which the agent is essentially never present and the ascent has no gradient.

\paragraph{Posterior agent.} The agent maintains a joint belief over the patch type and the unknown environmental parameter, on a discrete grid, updated in continuous time by the Poisson log-likelihood, subtracting $\lambda_\theta(k)\,\mathrm{d}t$ while waiting and adding $\log \lambda_\theta(k)$ at each encounter. It leaves when the predictive instantaneous rate $\mathbb{E}[\lambda_\theta(k)]$ falls below its estimate $\hat{R}$ of the opportunity cost, which is learned across patches by the delta rule $\hat{R} \leftarrow \hat{R} + \alpha(R_j - \hat{R})$ on the realized intake $R_j$ of patch $j$. The parameter belief is carried across patches by marginalizing over type at each departure. For panel D the grid is over $(\lambda_H, \lambda_L)$ with $\rho$ known, restricted to $\lambda_H > \lambda_L$ for identifiability; for panels E and F it is over $\rho$ with the rates known. In the binary case with $\rho$ known the departure time has a closed form, since the predictive rate is monotone in $u$.

\paragraph{Certainty equivalence.} Two variants share the machinery and differ only in taking maxima. The first replaces the parameter belief with its maximum likelihood value while retaining the type posterior; the second takes the joint maximum over parameter and type. Both then plan as if the point values were true. They are otherwise identical to the posterior agent, including the learned $\hat\rho$ and its initialization, so the comparison isolates the estimator.

\paragraph{Model-free learner.} The agent holds no belief about patches. It selects a giving-up time from a fixed grid, epsilon-greedy with exponentially annealed exploration, and updates that arm's value by the delta rule on the realized intake of the patch just left. Its state is a table indexed by policy with no referent in the environment.

\paragraph{Perturbation protocols.} Initialization sensitivity is swept in matched relative terms, the posterior agent starting at $\hat{R}_0 = R^\ast(1+e)$ and the model-free learner at arm values $q^\ast(1+e)$, where $q^\ast$ is the best attainable fixed giving-up intake, so both families are perturbed by the same fractional amount. Misspecification is imposed by giving the agent a uniform prior on $[0.85\,\rho_{\rm bel},\, 1.15\,\rho_{\rm bel}]$ with $\rho_{\rm bel} = \rho(1+m)$, which excludes the true $\rho$ once $|m| > 0.15$. The agent updates this belief from its encounters exactly as it updates its belief over patch type, so the failure at large $|m|$ is one of representation rather than of estimation. The model-free comparator is always its exact optimum, that is, its best case.

\paragraph{Numerics.} Inter-encounter times are drawn by inversion. Belief updates and departure checks for grid-based agents use a step of $0.25$ time units. A dwell time cap of $60$ time units prevents rules that can demand more chunks than a patch holds from running indefinitely; above the misspecification threshold the cap never binds and intake is identical at caps of $30$, $60$ and $120$, while below it only the depth of the collapse depends on the cap. Curves average $100$ realizations in panel D, $50$ in panel E and $200$ in panel F, and windowed intakes are computed from cumulative reward and time rather than from running averages.

\subsection*{Numerical methods}

Posteriors over $\lambda_1$ and over $(\lambda_H,\lambda_L)$ are evaluated on a grid in units of $\rho$, since depletion discretizes the rate to integer multiples of the decrement. The fraction posterior of Eq.~(\ref{softpost}) is a degree-$n$ polynomial in $p_H$ and is likewise evaluated on a grid. Mean learning curves, residence-time distributions and intake rates are estimated by Monte Carlo, with realization counts given in each figure caption. In the replenishing environment the deterministic orbit is iterated to the fixed point of Eq.~(\ref{dwelleq}) by bisection, and the lock-in band of Eq.~(\ref{rstar}) is evaluated from the flat-prior hypergeometric posterior over $N_H$. Eq.~(\ref{rstar}) evaluates the return at $\lambda^{\rm dep}=0$ rather than at the residual $R^*$ of Eq.~(\ref{reentry}), which keeps the $1/n_c$ scaling exact and places both edges above the thresholds an agent carrying the residual would use, leaving the band width unaffected.

\section*{Acknowledgments}
We acknowledge the Konstanz School of Collective Behavior, where important exchanges and work for this project occurred. AEH is supported by the DFG German Research Foundation (EXC 2117-422037984) and the Human Frontier Science Program (RGP006/2025). ZPK is supported by the U.S. National Science Foundation under awards DMS-2527338 and IIS-2616531.

\section*{Data availability}
This study generated no experimental data. All results derive from simulation and from the analytic expressions given in Methods; the scripts that produce every figure, together with their parameter settings and random seeds, are in the repository below.

\section*{Code availability}
All code used to generate the simulations, analyses, and figures reported in this study is available at \url{https://github.com/zpkilpat/learn-to-forage}.

\section*{Author contributions}
Conceptualization: Zachary P Kilpatrick, Ahmed El Hady \\
Formal analysis: Zachary P Kilpatrick \\
Funding acquisition: Zachary P Kilpatrick, Ahmed El Hady \\
Methodology: Zachary P Kilpatrick, Ahmed El Hady \\
Project administration: Zachary P Kilpatrick, Ahmed El Hady \\
Software: Zachary P Kilpatrick \\
Writing – original draft: Zachary P Kilpatrick, Ahmed El Hady \\
Writing – review \& editing: Zachary P Kilpatrick, Ahmed El Hady

%\nolinenumbers

\bibliographystyle{unsrt}
\bibliography{learning}

\pagebreak 

\setcounter{figure}{0}
\renewcommand{\thefigure}{S\arabic{figure}}
\section*{Supporting information}

\subsection*{S1 Text: Sensitivity to the prior}

\begin{figure}[b!]
\begin{center} \includegraphics[width=17cm]{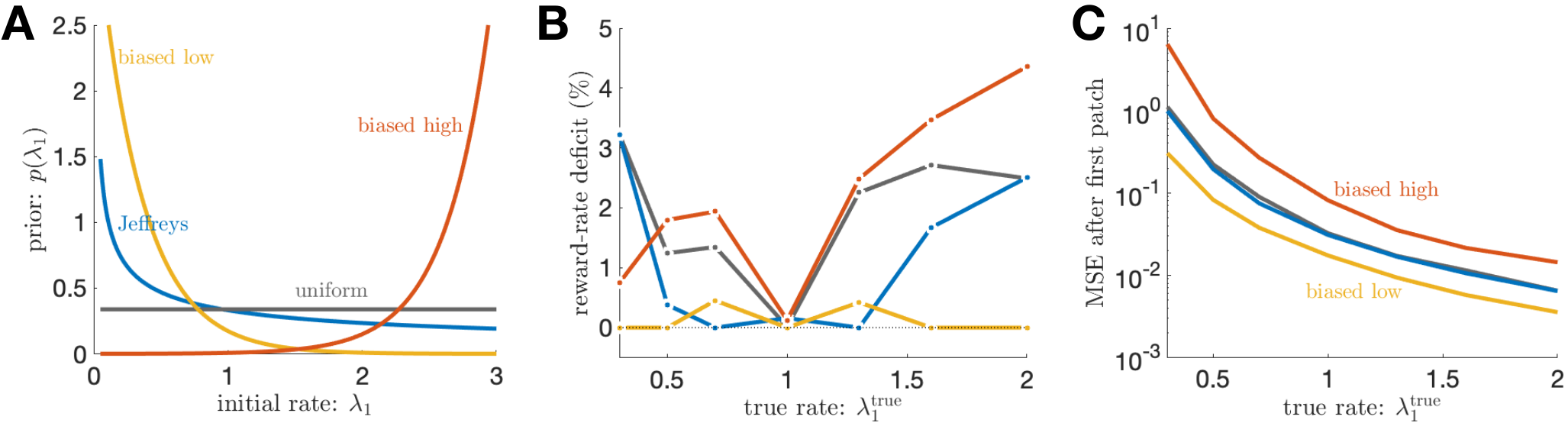} \end{center}
\caption{{\bf Foraging performance is insensitive to the prior even when early belief is not.} {\bf A.}~Four priors over the initial rate $\lambda_1$, the Jeffreys prior $\propto \sqrt{{\mc I}(\lambda_1; \rho)}$, a uniform prior, and two deliberately biased exponential priors favoring low and high rates, all normalized over the displayed interval. {\bf B.}~Shortfall in long-run reward rate against the best-performing prior at each true rate ($\rho = 0.1$, $\tau = 1$), for a forager using the learning-optimal policy of Section~\ref{sec:rates}, departing once its posterior-mean chunk count is consumed. No prior loses more than $4.4\%$ of intake anywhere. The curves are jagged and coincide at $\lambda_1^{\rm true} = 1$ because the policy is an integer chunk count, so a prior pays only where its rounded estimate lands on a different count than the best one. {\bf C.}~Normalized mean squared error of the rate estimate after the first patch departure, for the same four priors. Here they separate by a factor of $21$ in the poorest environment, narrowing to $4$ in the richest, with the biased-high prior worst and the biased-low best throughout.}
\label{figS1_priors}
\end{figure}

We used flat priors in the main text for tractability. Cognitive biases from movement orientation and spatial memory can strongly shape animals' foraging~\cite{beardsworth2021habitat,menzel2005honey,regular2013must,boyer2012non}, and enter a dynamical model as a pre-existing prior over the distribution of available resources~\cite{gil2009honeybees}. Other forms encode simple and objective assumptions about the variables of interest instead~\cite{kass1996selection}, among them the Jeffreys prior, which is invariant under changes of coordinates of the parameter vector~\cite{jaynes1968prior} and so attractive here since animals may be misestimating scaling parameters like patch size. Its form for the depleting process is derived in Methods, Eq.~(\ref{depoisfi}).

While the Jeffreys prior is improper, we can normalize it over a finite interval to compare it against alternatives (Fig.~\ref{figS1_priors}A). Both the Jeffreys and uniform priors peak at the low end of $\lambda_1$, and we additionally consider two strongly opinionated priors that favor low and high rates respectively. Foraging under all four is remarkably insensitive to the choice, since no prior costs more than $4.4$ percent of long-run reward rate at any true rate (Fig.~\ref{figS1_priors}B). This is not because the priors encode similar beliefs. After a single patch, the mean squared error of the rate estimate differs by a factor of $21$ across priors in the poorest environment, narrowing to $4$ in the richest, the biased-high faring worst and the biased-low best (Fig.~\ref{figS1_priors}C). Nor does accuracy determine intake, since the biased-low prior costs least in panel B for a reason unrelated to how well it estimates, underestimating $m_0$ making it depart earlier, which happens to sit closer to the reward-optimal count than consuming the patch does. Depletion supplies enough within-patch information to overwrite the prior within a few encounters, so the prior matters most where information is scarcest, in the poorest and earliest-visited patches, and is quickly overcome elsewhere. The same flatness in the reward surface returns in Section~\ref{sec:weakinstrument}, where the best rule carrying no model of the patches lands within half a percentage point of the posterior agent.

\subsection*{S2 Text: Joint inference of rates and composition}

\begin{figure}[t!]
\begin{center} \includegraphics[width=\textwidth]{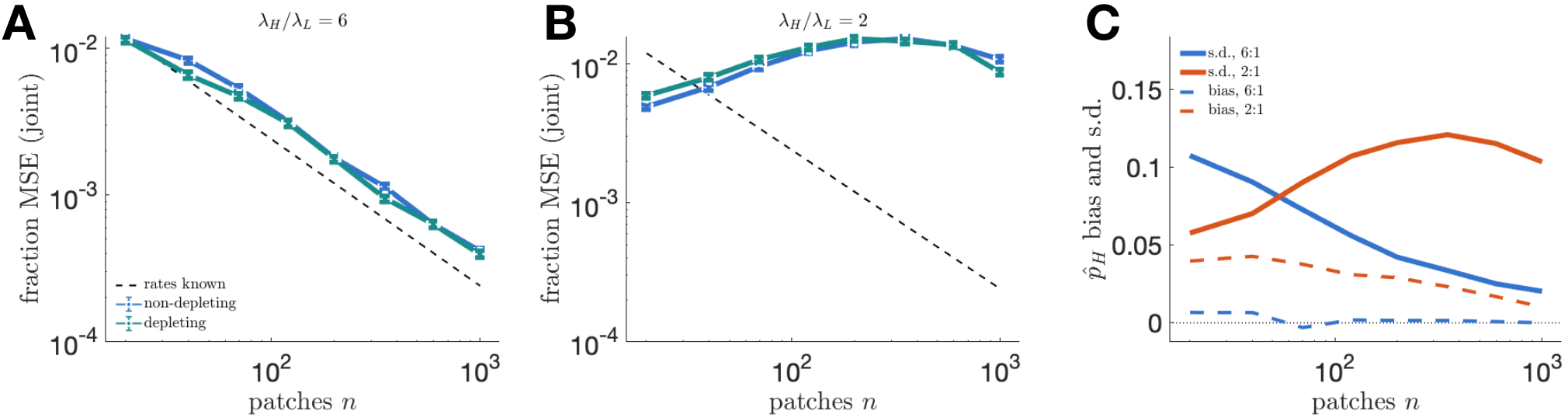} \end{center}
\vspace{-5mm}
\caption{\textbf{Under joint inference of the rates and the composition, depletion helps once the mixture is identified.} \textbf{A, B.} Composition MSE against number of patches $n$ under joint inference of $(\lambda_H,\lambda_L,p_H)$, non-depleting (blue) and depleting (teal), with Monte Carlo standard errors over $800$ realizations. Dashed is $p_H(1-p_H)/n$, the error attainable with the rates known. \textbf{A.} Well-separated rates, mean counts six versus one per visit. Both curves fall with fitted slopes near $-0.89$, and depletion is the better of the two at all eight sample sizes, by $8$ percent in geometric mean. \textbf{B.} Closely-spaced rates, mean counts four versus two. Both curves peak between $n=200$ and $350$ and then fall, and depletion's effect changes sign, costing $20$ percent at $n=20$ and decaying through the range. \textbf{C.} Bias (dashed) and standard deviation (solid) of $\hat{p}_H$ for the two separations, non-depleting. At the narrow separation the spread peaks near $n=350$ before falling, so the window of weak identification is finite.}
\label{figS2_joint}
\end{figure}

The main text shows that, with the rates known and each visit scored by the correct likelihood, composition learning falls as $1/n$ in patches and is unaffected by depletion (Fig.~\ref{fig3_composition}B). Depletion can act there only through the type readout, which it does not bias under the correct likelihood. When the agent must infer the rates and the composition jointly, a second and indirect channel opens, since the reduced encounter counts also change the rate estimates that feed the classification of each patch. Two effects compete. Depletion lowers the counts, which leaves less information per visit, and it makes the intervals within a visit informative about the initial rate, which sharpens what those counts imply. Which one shows depends on whether the mixture is identified at all.

Neither effect is large, so only aggregate statistics are trustworthy here. Two independent implementations of the same estimator, each at $800$ realizations, resolve individual sample sizes at different values of $n$, so we quote signs, geometric means and trends across the whole range rather than per-point significance.

{\em Where the rates are separable, depletion helps (Fig.~\ref{figS2_joint}A).} With well-separated rates both environments recover a decaying composition error, at fitted slopes near $-0.89$ against the $-1$ of the rates-known floor. The shortfall is expected rather than a failure of the estimator, since that floor is what a forager achieves knowing the rates and having only to count types, whereas a joint estimator pays for the rates too and so tracks the floor at a widening distance, a factor of $1.7$ by $n=1000$. Against that common backdrop the depleting estimate is the better of the two at all eight sample sizes, by eight percent in geometric mean, a sign test at $p = 0.004$, with no trend in $n$. The within-visit information that accelerates rate learning propagates upward, giving tighter rate posteriors, cleaner type assignments and a marginally better composition estimate.

{\em Where the rates are close, the same benefit appears once there is enough data to find it (Fig.~\ref{figS2_joint}B).} At a separation of two the picture is not stationary in $n$. Depletion costs twenty percent at $n=20$ and that cost decays through the range, crossing parity near $n=350$ and reaching a nominal nineteen percent advantage by $n=1000$, a trend of $-0.087$ in the slope of the log ratio against $\log n$. We do not lean on the reversal itself, which rests on the last point, but the decay is carried by all eight and by an independent implementation. The absolute error behaves the same way, rising to a peak between $n=200$ and $350$ before falling. At this separation the mixture is simply not identified until several hundred patches have been seen, and only then can the benefit that panel A shows immediately begin to express itself.

{\em The intervening regime is a finite window, not a pathology (Fig.~\ref{figS2_joint}C).} Decomposing the estimate shows what the peak is made of. At the narrow separation the bias falls from $0.040$ to $0.010$, so the estimator is consistent, while the spread of $\hat{p}_H$ first grows, from $0.058$ at $n=20$ to about $0.12$ near $n=350$, and then contracts to $0.104$. A flat prior over $p_H$ dominates a weak likelihood at small $n$ and holds the posterior mean near its center, so small samples look accurate for the wrong reason. As $n$ grows the likelihood takes over but concentrates along a ridge in which a lower $\lambda_H$ and a higher $p_H$ explain the data about as well as the truth, so the estimate wanders. By a few hundred patches the ridge is resolved and the spread falls again. The well-separated case skips this entirely, its bias at zero throughout and its spread falling from $0.108$ to $0.020$, the two spreads crossing near $n=70$.

The scope of the main-text claim is therefore unchanged. Composition learning is intrinsically slow because information about it arrives one patch at a time, and depletion does not change that at any separation, which is the result established with the rates known. Allowing the rates to be inferred as well leaves a modest benefit wherever the mixture can be identified at all.

\subsection*{S3 Text: Supporting analysis for the replenishment limit cycle}

\begin{figure}[t!]
\begin{center} \includegraphics[width=14cm]{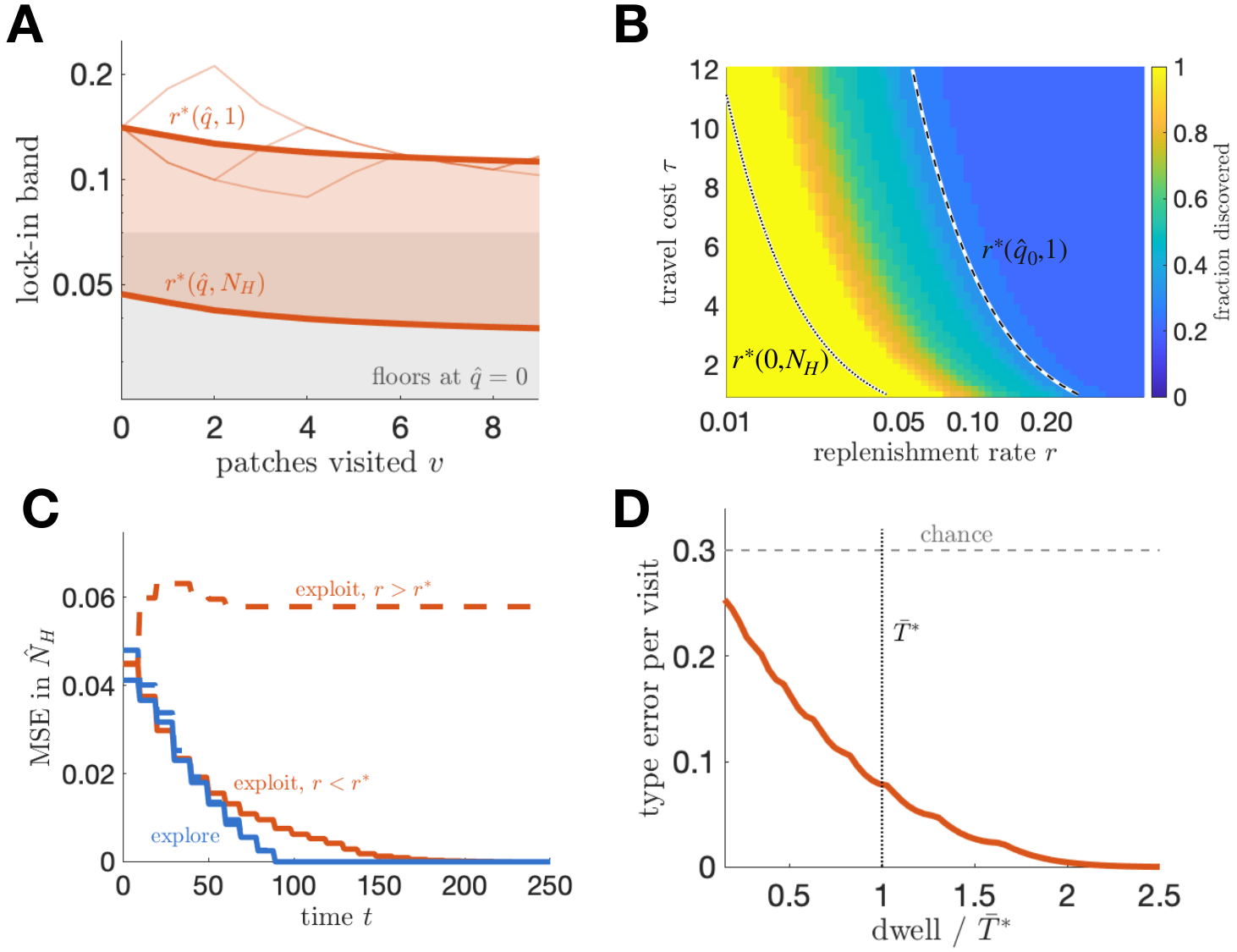} \end{center}
\vspace{-3mm}
\caption{\textbf{Lock-in, its dependence on travel cost, and the two things that limit composition learning.} \textbf{A.} The lock-in band of Eq.~(\ref{rstar}) descending as the agent's belief $\hat{q}(v,h) = (\mathbb{E}[N_H \mid v,h]-h)/(N-v)$ falls with experience, toward its floors at $\hat{q}=0$ (grey). The edges differ by exactly $N_H$, since $r^*(q,n_c) = r^*(q,1)/n_c$, so on the logarithmic axis only the band's position carries information. Faint traces are three individual sampling orders. \textbf{B.} Fraction discovered over the replenishment rate and travel cost, with both band edges overlaid. Costly transit lowers the whole band, since both edges scale inversely with $\bar{T}^*+\tau$, so lock-in sets in at lower recovery rates. \textbf{C.} Error in the posterior mean $\hat{N}_H$ against time. Below lock-in (solid) the information-seeker (blue) is no faster than the reward-seeker (red) while the estimate is coarse and pulls ahead as it tightens, reaching zero in about half the time; both resolve the composition exactly. Above lock-in (dashed) the reward-seeker plateaus at a floor set by the patches its cycle never reaches. \textbf{D.} Type-readout error per visit against dwell time, in units of the reward-optimal dwell time $\bar{T}^*$ (dotted), against the chance rate $N_H/N$ (grey).}
\label{figS3_replenish}
\end{figure}

{\em The phase of the orbit is only weakly corrected.} Equation~(\ref{dwelleq}) fixes the dwell time but not the phase, so a schedule in which the high patches were visited in quick succession is consistent with the same dwell time and a different, clumped orbit. Such a schedule is not itself an attractor, since a patch visited longer ago is more recovered, so the greedy return to the most recovered patch preferentially picks up whichever patch has fallen behind, and the schedule is pulled back toward equal spacing. The correction is weak, however. With the agent's opportunity cost held at its fixed-point value $R^*$, as Section~\ref{sec:replenish} specifies, the leading multiplier of the return map in the phase direction is about $0.98$ at $r = r^*$, falling to $0.95$ at $4r^*$, so a phase perturbation decays over tens of cycles rather than a few. The relaxation in Fig.~\ref{fig5_replenish}C is faster because that simulation also updates its estimate of the long-run average from its own intake history, and an agent correcting its own opportunity cost re-equalizes the schedule in about three cycles, whereas with the estimate fixed the same perturbation takes far longer. The recovery direction behaves oppositely, since Eq.~(\ref{reentry}) gives $\lambda^{\rm dep} = R^*$ whatever the arrival rate, so that coordinate contracts to its fixed point in a single visit and displacement away from the patch set is forgiven within one cycle.

{\em The band descends with belief (Fig.~\ref{figS3_replenish}A).} The threshold of Eq.~(\ref{rstar}) depends on the agent's belief $\hat{q}$ that an unvisited patch is high, and that belief falls as the agent types patches and finds the high ones. The band therefore descends through the bout, from its prior position toward the floors reached when the agent is certain nothing high remains. Its two edges are not independent, since $r^*(q,n_c)$ is inversely proportional to $n_c$, so the band is a window of fixed width $N_H$ whose position alone changes. The descent is partial on average, since in an arena of ten patches with three high ones the agent rarely becomes certain within the bout, and a floor is reached only in those realizations where the last high patch has been located. This is what makes the discovery boundary of Fig.~\ref{fig5_replenish}D a band rather than a line.

We take the belief from the flat-prior hypergeometric posterior over $N_H$, so that one belief serves the whole analysis. The alternative of giving the agent the true $N_H$ and letting it infer only which patches are high is tempting for tractability but it makes the expected belief exactly $N_H/N$ at every $v$, so no descent is possible, and it diverges in realizations where every remaining unvisited patch is high.

{\em Travel cost shifts the band (Fig.~\ref{figS3_replenish}B).} Both edges scale inversely with $\bar{T}^*+\tau$, so costly transit lowers the whole band. When travel is expensive an agent will accept a less recovered patch rather than pay to reach an unknown one, so lock-in sets in at lower recovery rates and the discovered fraction falls sooner in $r$.

{\em Information-seeking resolves the composition sooner, but not sooner throughout (Fig.~\ref{figS3_replenish}C).} Below lock-in the information-seeker is initially the slower of the two, since it spends early visits on untyped patches while the reward-seeker is also dragged outward and types patches incidentally. As the accuracy demand tightens the ordering reverses, and the information-seeker reaches one percent of its initial error in half the reward-seeker's time, though the reward-seeker eventually gets there too. Above lock-in the reward-seeker's error plateaus instead, at a floor set by the patches its cycle never reaches, so the advantage becomes permanent rather than transient.

{\em Sorting is a property of value-sensitive return.} The steady cycle is composed almost entirely of high patches, and two candidate explanations must be separated. It could be that recovery dynamics alone select the high patches, or that the agent's value comparison does. Recovery fraction is type-independent, so an agent returning to whichever patch has recovered the largest fraction of its own cap visits high patches at their base rate $N_H/N$, which is $0.30$ here. The reward-seeker instead holds purity at one for every rate gap below about $0.7$, falling to $0.44$ only at a gap of $0.92$ where a fully recovered low can beat a partially recovered high. Sorting is therefore done by the value comparison and not by the recovery process, and it survives until the two types are nearly indistinguishable.

{\em Stochastic sampling slows discovery without dissolving the orbit.} With stochastic encounters, the system has an invariant distribution concentrated near the deterministic orbit rather than a periodic orbit proper. The departure condition becomes a threshold on a discrete count, so the agent leaves after $K^* = \lceil (\lambda^{\rm in} - R^*)/\rho \rceil$ chunks and its dwell time is a sum of independent exponentials, with a mean that sits systematically below $\bar{T}^*$ because the count rounds up while the continuous dwell time does not (Fig.~\ref{fig5_replenish}F). Discovery slows for a separate reason. A visit of duration $T$ to a patch of initial count $m$ yields a count drawn from ${\rm Binomial}(m, 1-e^{-\rho T})$, and the maximum a posteriori readout has an error rate available in closed form. At the reward-optimal dwell time $\bar{T}^*$, that rate is $0.078$, so roughly one patch in thirteen is misassigned, and at a quarter of that dwell time it rises to $0.224$, against a chance rate of $0.30$ (Fig.~\ref{figS3_replenish}D). Since every misread patch must be revisited before the composition resolves, this is the mechanism behind the slower discovery transient in the stochastic case. It also determines how short a visit can be before the agent stops learning the map altogether. The degradation arises from measurement noise in the type readout rather than from depletion itself, since as in the binary environment a correctly specified agent learns composition no more slowly under depletion than without it once visits are long enough to resolve patch type.

\subsection*{S4 Text: Discriminability and the number of patch types}

The main text fixes $N_T = 2$, though the model of Section~\ref{sec:rates} is written for any finite $N_T$. Two questions follow, how the separation between rates governs the value of moving, and how much a forager loses by carrying a model of the environment coarser than the environment itself.

\begin{figure}[t!]
\begin{center} \includegraphics[width=17cm]{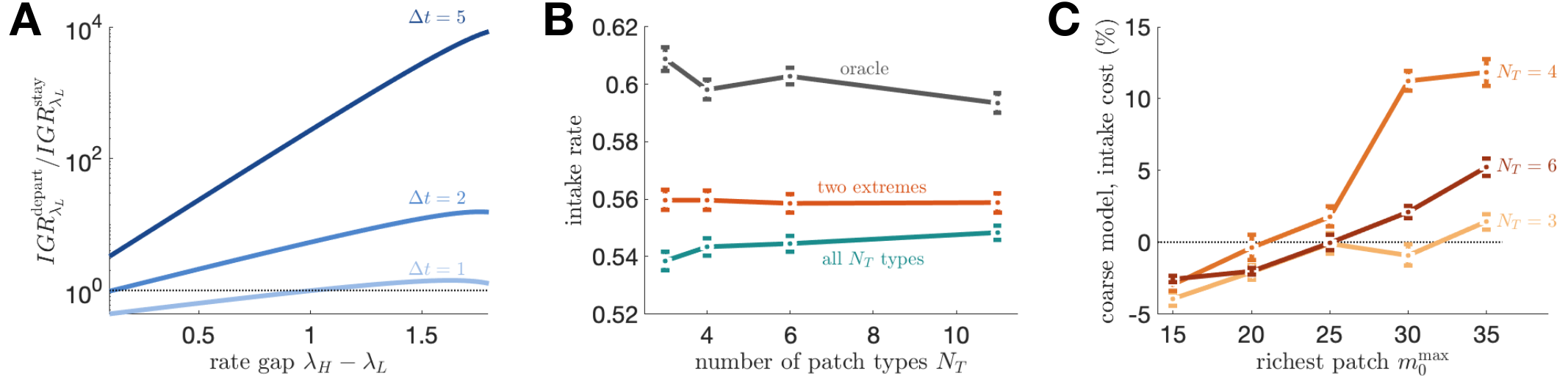} \end{center}
\vspace{-3mm}
\caption{\textbf{A coarse model class resolves patch type faster than a correct one, and whether that pays is set by the spread of richness it has to cover.} \textbf{A.} The stay-versus-go information ratio of Eq.~(\ref{igrstay}) against the rate gap, at three time horizons $\Delta t$, with $\lambda_H = 2$ and $\tau = 2$. The ratio is exponential in the gap, since over an increment of one time unit departing overtakes staying only once the rates differ by about half of $\lambda_H$, while over two or more it wins at every gap shown. \textbf{B.} Long-run intake against the number of patch types $N_T$, with richness held in $m_0 \in [5,15]$ so only the number of types varies. Grey, an oracle that knows each patch's richness on entry; teal, an agent that models all $N_T$ types; red, one that models only $m_0^{\min}$ and $m_0^{\max}$. \textbf{C.} The coarse agent's intake cost relative to the full model, positive meaning the coarse agent loses, against the richest patch present, at three values of $N_T$. Error bars are standard errors over six seeds; $\rho = 0.1$, $\tau = 2$ throughout.}
\label{figS4_types}
\end{figure}

{\em Discriminability sets the value of departure} (Fig.~\ref{figS4_types}A). Section~\ref{sec:binary} gives the information gain rate about an under-sampled rate as $O(p_L)$ from departing and $O(p_L e^{-(\lambda_H-\lambda_L)\Delta t})$ from staying, so their ratio is exponential in the gap. Plotting it fixes the scale. Over an increment of one time unit, departing overtakes staying only once the rates differ by about half of $\lambda_H$, while over two units or more it wins at every separation, since the chance that the occupied patch is of the other type decays with each further interval observed. Closely spaced rates therefore hold a forager in place both because each visit is less diagnostic of type, as the effective sample size of Eq.~(\ref{neff}) records, and also because the option value of leaving is smaller.

{\em A coarse model class is misspecified but decisive} (Fig.~\ref{figS4_types}B). We compare an agent that models all $N_T$ richness values against one that models only $m_0^{\min}$ and $m_0^{\max}$ and assigns every patch to whichever of them the evidence favors, both using the plug-in departure rule of Section~\ref{sec:rewardinfo}, and both against an oracle that knows each patch's richness on entry. Over $N_T$ from three to eleven at a threefold richness range, every model class we tried lands between $88$ and $94$ percent of the oracle, a spread narrower than the gap to the oracle itself. The cost of modeling the environment coarsely is smaller than the cost of not knowing the occupied patch's type.

More than that, the coarse agent is the better of the two at every $N_T$ over this range, by up to four percent, despite being wrong about every intermediate type by construction. The two agents fail differently. Under a plug-in rule the dominant error is overstaying a poor patch while the posterior is still spread across neighboring hypotheses, and the coarse model's two hypotheses are far apart, so each observation is strongly diagnostic and the agent commits early. Against that it misassigns every type it does not carry, to whichever extreme is nearer.

{\em Which effect wins is set by the richness range} (Fig.~\ref{figS4_types}C). Widening the spread from a threefold to a sevenfold ratio, holding $N_T$ fixed, moves the coarse agent from winning by two to four percent to losing, with the sign changing somewhere between a fourfold and a sevenfold ratio depending on $N_T$. A patch assigned to the wrong extreme is then further from its own optimum, and the misassignment cost overtakes the decisiveness it buys. The number of types matters much less, since below the crossover the three curves lie within about two percent of one another. One case departs from this, $N_T = 4$ at the two widest ranges, where the omitted types sit near the likelihood midpoint between the extremes and the coarse posterior is maximally ambiguous, and there the cost reaches twelve percent, against at most five percent at neighboring $N_T$.

The reversal is possible only because both agents are heuristics. An agent carrying the correct model and playing the optimal policy could not be beaten by a misspecified one, and the coarse model wins only against a rule hurt more by slow type resolution than by wrong hypotheses. Nor does changing that rule repair it, since averaging the departure policy over the posterior, rather than rounding the posterior mean and mapping that to a policy, leaves the ordering intact and is itself worse at larger $N_T$. Neither is optimal, and they are competing heuristics.

Committing early to a coarse description buys intake in three places here. In S1~Text the biased-low prior costs least of the four compared, for a reason unrelated to how well it estimates, and in Section~\ref{sec:weakinstrument} the best rule carrying no model of the patches comes within half a percentage point of the posterior agent. Here a model class wrong about most of the environment outperforms one right about all of it, provided the environment is not too spread out. In each case accuracy and intake come apart, and in each the advantage is exactly what fails when the description has to change.
\end{document}